\documentclass[fleqn,usenatbib]{mnras}
\usepackage{newtxtext,newtxmath}
\usepackage{amssymb}
\usepackage{mathrsfs}
\usepackage{xspace}

\usepackage{fix-cm}
\usepackage[T1]{fontenc}
\usepackage{ae,aecompl}

\newcommand{\rsun}{\,\mbox{$\rm R_{\odot}$}\xspace}

\newcommand{\msun}{\,\mbox{$\rm M_{\odot}$}\xspace}

\newcommand{\ionjb}[2]{#1{~\sc{\romannumeral #2}}}

\newcommand{\kms}{\hbox{km~s$^{-1}$}\xspace}

\newcommand{\vsini}{\hbox{$v$\,sin\,$i$}\xspace}

\newcommand{\degs}{$\degr$\xspace}
\newcommand{\chisq}{$\chi^{2}$\xspace} 
\newcommand{\chisqr}{$\chi_r^{2}$\xspace}

\newcommand{\ha}{H$\alpha$\xspace}

\newcommand{\gjsn}{\hbox{GJ 791.2A}\xspace}

\newcommand{\gjsfa}{\hbox{GJ 65A}\xspace}
\newcommand{\gjsfb}{\hbox{GJ 65B}\xspace}

\newcommand{\iphot}{$I^c_{\rm phot}$\xspace}
\newcommand{\ispot}{$I^c_{\rm spot}$\xspace}

\newcommand{\tphot}{$T_{\rm phot}$\xspace}
\newcommand{\tspot}{$T_{\rm spot}$\xspace}

\usepackage{graphicx}	% Including figure files
\usepackage{amsmath}	% Advanced maths commands
\usepackage{amssymb}	% Extra maths symbols

\title[Starspots and differential rotation on GJ 65AB]{Surprisingly different starspot distributions on the near equal-mass equal-rotation-rate stars in the M dwarf binary GJ 65 AB}

\author[J.R.~Barnes et al.]
{J.R.~Barnes et al. $^{1}$ \\
$^{1}$ School of Physical Sciences, The Open University, Walton Hall, Milton Keynes MK7 6AA, UK. \\
}

\author[J.R.~Barnes et al.]
{J.R.~Barnes$^{1}$, 
S.V.~Jeffers$^{2}$,
C.A.~Haswell$^{1}$,
H.R.A.~Jones$^{3}$,
D.~Shulyak$^{2}$, \newauthor
Ya.V.~Pavlenko$^{4}$,
J.S.~Jenkins$^{5}$
\\
$^{1}$ School of Physical Sciences, The Open University, Walton Hall, Milton Keynes MK7 6AA, UK. \\
$^{2}$ Institut f\"{u}r Astrophysik, Georg-August-Universit\"{a}t, Friedrich-Hund-Platz 1, Friedrich-Hund-Platz 1, D-37077 G\"{o}ttingen. Germany. \\
$^{3}$ Centre for Astrophysics Research, University of Hertfordshire, College Lane, Hatfield AL10 9AB, UK \\
$^{4}$ Main Astronomical Observatory of the National Academy of Sciences of Ukraine, Golosiiv Woods, Kyiv-127, 03680, Ukraine. \\
$^{5}$ Departamento de Astronom\'{i}a, Universidad de Chile, Camino del Observatorio 1515, Las Condes, Santiago. Chile. \\
}

\date{Accepted 2017 June 12.}

\pubyear{2017}

\begin{document}

\label{firstpage}
\pagerange{\pageref{firstpage}--\pageref{lastpage}}
\maketitle

% Abstract of the paper
\begin{abstract}

We aim to understand how stellar parameters such as mass and rotation impact the distribution of starspots on the stellar surface. To this purpose, we have used Doppler imaging to reconstruct the surface brightness distributions of three fully convective M dwarfs with similar rotation rates. We secured high cadence spectral time series observations of the $5.5$ AU separation binary GJ 65, comprising \gjsfa (M5.5V, $P_{\rm rot} = 0.24$\ d) and \gjsfb (M6V, $P_{\rm rot} = 0.23$\ d). We also present new observations of \gjsn (M4.5V, $P_{\rm rot} = 0.31$\ d). Observations of each star were made on two nights with {\sc uves}, covering a wavelength range from $0.64\,-\,$\hbox{$1.03$ \micron}. The time series spectra reveal multiple line distortions, which we interpret as cool starspots and which are persistent on both nights suggesting stability on the timescale of 3 days. Spots are recovered with resolutions down to $8.3$\degs at the equator. The global spot distributions for \gjsn are similar to observations made a year earlier. Similar high latitude and circumpolar spot structure is seen on \gjsn and \gjsfa. However, they are surprisingly absent on GJ 65B, which instead reveals more extensive, larger, spots concentrated  at intermediate latitudes. All three stars show small amplitude latitude-dependent rotation that is consistent with solid body rotation. We compare our  measurements of differential rotation with previous Doppler imaging studies and discuss the results in the wider context of other observational estimates and recent theoretical predictions.

\end{abstract}


%%%%%%%%%%%%%%%%%%%%%%%%%%%%%%%%%%%%%%%%%%%%%%%%%%

%%%%%%%%%%%%%%%%% BODY OF PAPER %%%%%%%%%%%%%%%%%%

%%%%%%%%%%%%%%%%%%%%%%%%%%%%%%%%%%%%%%%%%%%%%%%%%%%%%%%%%%%%%%%%%%%%%%%%%%%%%%% INTRODUCTION %%%%%%%%%%%%%%%%%%%%%%%%%%%%%%%%%%%%%%%%%%%%%%%%%

\section{Introduction}
\protect\label{section:intro}
%Starspots have been studied in detail for a number of stars using Doppler imaging techniques \citep{strassmeier09starspots}. 
%Many G and K dwarfs have been mapped using Doppler imaging techniques \citep{strassmeier09starspots}, 
The surface brightness distributions of many G and K dwarfs have been reconstructed using Doppler imaging techniques \citep{strassmeier09starspots},
but there are very few brightness images of M dwarfs, largely due to their intrinsic faintness. In rapidly rotating G and K stars, a solar-like dynamo mechanism under the action of rapid rotation \citep{moreno92fluxtubes,schussler96buoy} has been inferred from spot patterns (e.g. \citealt{barnes98aper,jeffers07abdor,marsden06}). The simultaneous presence of low latitude spots { implies} that distributed dynamo activity is { also} present. This possibility has been investigated by \citet{brandenburg05dynamo} from a theoretical perspective in light of helioseismology findings. { The first images of early-M dwarfs revealed that spots are distributed relatively uniformly in longitude and latitude \citep{barnes01mdwarfs,barnes04hkaqr}, with no evidence for the strong polar spots seen in earlier spectral types.}

\begin{table*}
\begin{tabular}{lcccccc}
\hline
~~~~~~~~~Star      & SpT   & Vmag   & Exp & S/N          & S/N             & Number of spectra \\
		   &       &        & [s] & (extracted)  & (deconvolved)   & observed / used \\
\hline
GJ 791.2A (HU Del) & M4.5V & 13.13  & 180 & 56  $\pm$ 6  & 3120 $\pm$  316 & 163 / 163 \\
GJ 65A (BL Cet)    & M5.5V & 12.7   & 180 & 105 $\pm$ 10 & 5970 $\pm$  528 & 178 / 178 \\
GJ 65B (UV Cet)    & M6V   & 13.2   & 180 & 91  $\pm$ 10 & 5076 $\pm$  526 & 178 / 154 \\
\hline
\end{tabular}
\vskip 2mm
\caption{Summary of properties and observations made with {\sc vlt}/{\sc uves} on 2015 September 25 \& 28. The first 24 observations of \gjsfb were severely affected by a flare and were not use for imaging. A flare was seen on \gjsn at the end of the September 25 time series.}
\protect\label{tab:targets}
\end{table*}

%HU Del (GJ 791.2A)  	& M4.5V 	&   &   &   &   &   & 13.0 & 180    & 55.6  $\pm$ 5.7 & 3119.6 $\pm$  315.9   & 163  \\
%BL Cet (GJ 65A)   	& M5.5V 	&   &   &   &   &   & 12.7 & 180    & 105.1 $\pm$ 9.8 & 5969.7 $\pm$  528.3   & 178  \\
%UV Cet (GJ 65B)   	& M6V		&   &   &   &   &   & 13.2 & 180    & 90.8  $\pm$ 9.6 & 5076.3 $\pm$  525.7   & 178  \\

{ As stars become fully convective, at spectral type M3.5V, distributed dynamo activity is expected to be the sole mechanism by which magnetic fields can be generated and sustained. %The amplification and maintenance of magnetic fields in fully convective stars must result predominantly from distributed dynamo activity.
}
%More recent observations of fully convective stars reveal a similar picture.
Brightness images of only three fully convective mid-M stars have been published: V374 Peg \citep{morin08v374peg} and G 164-31 \citep{phanbao09} are both M4V stars, while \gjsn\ (HU Del) is an M4.5V star (\citealt{barnes15mdwarfsB15}; hereafter B15). V374 Peg reveals weak spots at intermediate latitudes. Little coherence of spot patterns was seen from observations made a night apart, though the moderate S/N ratio and phase coverage may have contributed to a lack of consistency between the image reconstructions. In contrast to V374 Peg, G 164-31 revealed only polar filling, but no low or intermediate spots, despite observations with good S/N ratio. A map of \gjsn\ resolved numerous spots with high latitude circumpolar structure, and spots concentrated at low-latitudes and distributed at all phases or longitudes. Two-temperature modelling was used by B15, requiring low contrast spots with \hbox{\tphot - \tspot} \hbox{= 300 K} (derived from model atmospheres). B15 also interpreted line profile variations in the late-M star, LP 944-20 (M9V), as cool spots and recovered only high latitude spots using low spot/photosphere contrasts of \hbox{\tphot - \tspot} \hbox{= 100 - 200 K}, confirming the earlier trend of decreasing spot contrast with decreasing photospheric temperature noted by \citet{berdyugina05starspots}.

Despite the expectation of distributed dynamo activity, Zeeman broadening of absorption lines using unpolarised spectra has been used to infer large magnetic fields of \hbox{2\,-\,4 kG} (\citealt{saar85adleo}; \citealt{johnskrull96mdwarfs}) { on M3.5 and M4.5V stars}. The large scale magnetic field topology of stars can be studied in more detail using polarised Stokes V time series observations. { Large scale fields with preferentially toroidal
and {\em non-axisymmetric} poloidal configurations are found in the case of M0V\,-\,M3V stars \citep{donati08mdwarfs}, while {\em axisymmetric} large-scale poloidal fields are found at the M4V fully convective boundary \citep{morin08mdwarfs}.} For M5\ -\ M8V stars, firmly in the fully convective regime, some stars exhibit strongly axisymmetric dipolar fields while others show weak fields with a significant non-axisymmetric component \citep{morin10mdwarfs}.  Simulations by \citet{gastine12} have subsequently found a bifurcation of the magnetic field geometry. This was investigated in the context of M dwarfs by \citet{gastine13}, who found that the dynamo bistability is most pronounced for stars with small Rosby numbers, resulting in either a dipolar field or multipolar field configuration.
%, $R_{\rm o} = P_{\rm rot}/\tau_{\rm c}$ (where $P_{\rm rot}$ and $\tau_{\rm c}$ are the rotation period and convective turnover time). 
Stronger magnetic fields lead to dipolar field geometry while weaker magnetic fields give rise to multipolar geometry. 
%The dichotomy of field configurations may manifest in brightness images as the presence or absence of a polar spot or circumpolar structure. One might expect that a predominantly axisymmetric simple dipolar field would result in suppression of convection leading to cool polar spots, as is often seen in rapidly rotating G and K stars; and the simulations of \citet{gastine13} and \citet{yadav15} lead to magnetic flux predominantly in the polar regions. 
%Both \gjsn (M4.5V) and \hbox{G 164-31} (M4V) exhibit polar spot structure, while \hbox{V374 Peg} (M4V) shows no evidence for polar spots. 
{ Although they show different starspot patterns in their brightness images, G 164-31 and V374 Peg both show axisymmetric dipolar magnetic fields with one polarity predominantly in the polar region. On the other hand, the almost identical components of GJ 65, reveal magnetic field structure that is axisymmetric in the case of \gjsfb\ (M6V) and non-axisymmetric for \gjsfa\ (M5.5V) modes. 
%The apparent inconsistencies between brightness and magnetic images indicates that comparisons may not be straightforward.
}

%Despite nearly identical masses and rotation rates, the secondary exhibits an axisymmetric, dipolar-likeglobal field with an average strength of 1.3kG while the primary has a much weaker, more complex, and non-axisymmetric 0.3kG field.

%Fully convective stars are not expected to exhibit significant latitude dependent rotation at the photosphere since this phenomenon is largely the result of the tachocline in FGK and early M stars. 

{  Differential rotation arises as a result of convection in the presence of rotation (due to Coriolis forces). Convection zone depth and stellar rotation rate might thus be considered important parameters governing its magnitude.
Following the first measurement of differential rotation using Doppler images from closely separated epochs \citep{donati97zdi},
a parameterised solar-like differential rotation model was incorporated directly into the line modelling and image recovery process by \citet{petit02} and subsequently by \citet{barnes05diffrot}. \citet{reiners03diffrot} also used Fourier transform techniques to study absorption line profile morphology and found that F dwarfs possess even higher degrees of differential rotation than G dwarfs. Subsequent work, specifically on G and K dwarf stars, by a number of authors using Doppler imaging methods (\citealt{jeffers07abdor,marsden06hd171488} and \citealt{marsden11hd141943}) have added to the sample of stars with differential rotation measurements. While F and G stars with a relatively small convection zone were found to exhibit the strongest differential rotation, by early M spectral type, the differential rotation was consistent with solid body rotation (i.e. no latitude dependent rotation) within the measurement uncertainties \citep{barnes04hkaqr,barnes05diffrot}. Further measurements of differential rotation on M dwarfs using the sheared image technique have also been made by \citet{donati08mdwarfs} who find significant differential rotation for early-M dwarfs (contrary to the results reported by \citealt{barnes05diffrot}) and by \cite{morin08mdwarfs} who find differential rotation rates for mid-M dwarfs that are typically ten times smaller.}

%and \citet{morin08v374peg,morin08mdwarfs}. { Early-M dwarfs are found to possess significant differential rotation, whereas Mid-M dwarfs show much smaller differential rotation.

%with $\Delta\Omega/\Omega \sim 5$ per cent.

 %For rapidly rotating stars, as stellar mass decreases and the relative size of the convection zone increases, the surface differential rotation becomes weaker \citep{barnes05diffrot}. 
{ The variation in differential rotation with spectral type and rotation rate has also been modelled using mean field hydrodynamics by \cite{kueker11diffrot} and \cite{kitchatinov12}. \cite{browning08} found that magnetic fields strongly quench the differential rotation. \citet{gastine13} predict that stars in the dipolar field branch should yield very weak differential rotation, while multipolar field configurations might in fact allow significant differential rotation. \cite{yadav15} similarly investigated a fully convective stellar model with differential rotation reduced by a strong magnetic dipolar field orientated with the rotation axis. Thus while significant non-solid body rotation is possible in fully convective stars, those stars that are more rapid rotators (with smaller Rosby numbers) and therefore more magnetically active, are likely to exhibit the lowest differential rotation rates.
}

Here we present Doppler images of three fully convective stars, including new brightness maps of \gjsn following our image derived a year earlier (B15). While \gjsn and \gjsfa, both show similar starspot patterns, our image of \gjsfb shows shows a markedly different spot distribution and a greater degree of spot filling. A brief introduction to the individual targets is presented in \S \ref{section:intro_targets} followed by a description in \S \ref{section:observations} of the observations and techniques used to derive the Doppler images. Images and differential rotation measurements are presented in \S \ref{section:results} with further discussion and concluding remarks in \S \ref{section:discussion}.

%%%%%%%%%%%%%%%%%%%%%%%%%%%%%%%%%%%%%%%%%%%%%%%%%%%%%%%%%%%%%%%%%%%%%%%%%%%%%%% TARGETS %%%%%%%%%%%%%%%%%%%%%%%%%%%%%%%%%%%%%%%%%%%%%%%%%

\section{Targets}
\protect\label{section:intro_targets}

The three targets in this study are nearby, bright, fully convective M4.5\,-\,M6 dwarfs, with ${\rm V} = 12.7 - 13.2$ (Table \ref{tab:targets}). They are relatively young and exhibit rapid rotation, making them suitable objects for brightness Doppler imaging.

\subsection{GJ 791.2A}
\protect\label{section:intro_gj791}

GJ 791.2AB is a nearby unresolved astrometric binary with apparent magnitudes of ${\rm V} = 13.13$ and ${\rm I} = 9.97$ \citep{hosey15}. { Based on kinematics, \citet{montes01members} do not consider it to be a member of the 0.6 Gyr Hyades Supercluster. More recently, \citet{benedict16} find that the components lie close to the \hbox{0.1 Gyr} model of \citet{baraffe15}.} The astrometric orbit of the unresolved GJ 791.2AB system determined by \citet{benedict00gj791} has been re-analysed by Benedict et al. 2016 who find an orbit with $P = 538.59$ d. GJ 791.2B is 3.27 magnitudes fainter (20.3 times smaller flux) in the V-band, though as in B15, we do not see the secondary component in the photospheric absorption lines. The Benedict et al. 2016 respective component masses for \gjsn and GJ 791.2B of $M_A = 0.237 \pm 0.004$\msun and $M_B = 0.114 \pm 0.002$\msun are slightly lower than the earlier Benedict et al. 2000 estimates. B15 found a rotation period of $P_{\rm rot} = 0.3088$ d and obtained Doppler images revealing numerous low contrast spots. Despite exhibiting significant activity, \citet{hosey15} find only 8.2 mmag variability in the $I$ band (from 148 observations spanning 30 nights over a 7.25 yr time span), in common with other targets of similar spectral type, as might be expected from a star with low contrast spots distributed across its surface (B15).

\subsection{GJ 65A and GJ 65B}
\protect\label{section:intro_gj65}

GJ 65 (Luyten 726-8) was first reported as a new binary with a large proper motion by \citet{luyten49gj65} who also identified flaring activity on the fainter component, GJ 65B (UV Cet). Subsequent flaring activity was reported by a number of authors, while \citet{bopp73gj65} considered that since both components, \gjsfa (BL Ceti) and \gjsfb are of similar spectral type (M5.5V and M6V respectively), they should both be treated as flare stars. Photometry and spectroscopy by \citet{bopp73gj65} enabled detailed study of flaring events on the unresolved pair. UV Cet has become the class prototype of stars that undergo rapid photometric brightening due to dramatic flaring activity.

GJ 65 is a visual binary, which at a distance of only 2.68 pc \citep{henry97recons}\footnote{http://recons.org} is the 6th closest stellar system to the Sun, with unresolved magnitudes of ${\rm V} = 12.08$  \citep{zacharias12} and  ${\rm I} = 8.93$ \citep{denis03}. \citet{mason01catalogue} give respective magnitudes for \gjsfa and \gjsfb of V = 12.70 and 13.20. \citet{montes01members} found that GJ 65 is a possible member of the 600 Myr Hyades supercluster moving group. More recently, \citet{kervella16gj65} discussed the age and population membership of \hbox{GJ 65} in detail, showing it to be consistent with a 200\,-\,300 Myr old thin disk system. \citet{benedict16} find $M_A = 0.120 \pm 0.003$\msun and $M_B = 0.117 \pm 0.003$\msun. These masses are in agreement with those given by \citet{kervella16gj65} who also give respective projected rotation velocities of \vsini = \hbox{$28.2 \pm 2$ \kms} and \hbox{$30.6 \pm 2$ \kms}. The binary orbital period of \hbox{$P_{\rm bin} =$}\ \hbox{$26.284$ yrs} \citep{kervella16gj65} implies a semi-major axis of 5.5 AU, while the astrometric eccentricity, $e = 0.619$. Because the orbital separation is large compared with the stellar radii, tidal effects will be small and they are unlikely to undergo the increased activity levels commonly observed in RS CVn binaries. The projected separation of \gjsfa and \gjsfb in 2015 September was \hbox{2.18 arcsec}, close to the maximum separation of \hbox{2.19 arcsec}. { Magnetic maps of both \gjsfa and \gjsfb have been derived by \cite{kochulov17}, who find different global magnetic field topologies.}
%enabling simultaneous observations of both stars in good seeing conditions.

\begin{table*}
\begin{tabular}{lcccc}
\hline
~~~~~~~~~Star      & $i$         & \vsini          & P                    & $\Delta\Omega$    \\
		   & [degs]      & [\kms]          & [d]                  & [rad d$^{-1}$]    \\
\hline
GJ 791.2A (HU Del) & $55 \pm 4$  & $35.1 \pm 0.2$  & $0.3085 \pm 0.0005$ & $0.035 \pm 0.002$ \\
GJ 65A (BL Cet)    & $60 \pm 6$  & $28.6 \pm 0.2$  & $0.2430 \pm 0.0005$ & $0.031 \pm 0.054$ \\
GJ 65B (UV Cet)    & $64 \pm 7$  & $32.2 \pm 0.2$  & $0.2268 \pm 0.0003$ & $0.026 \pm 0.040$ \\
\hline
\end{tabular}
\vskip 2mm
\caption{Target properties derived from \chisq minimisation with DoTS.}
\protect\label{tab:properties}
\end{table*}

%%%%%%%%%%%%%%%%%%%%%%%%%%%%%%%%%%%%%%%%%%%%%%%%%%%%%%%%%%%%%%%%%%%%%%%%%%%%%%% OBSERVATIONS %%%%%%%%%%%%%%%%%%%%%%%%%%%%%%%%%%%%%%%%%%%%%%%%%

\section{Observations and Analysis methods}
\protect\label{section:observations}

GJ 791.2A, \gjsfa and \gjsfb  were observed on the nights of 2015 September 25/26 and 28/29 at the Very Large Telescope ({\sc vlt}) with the Ultraviolet and Visual \'{E}chelle Spectrograph ({\sc uves})\footnote{ESO programme 095.D-0291(A)}. The observations were made with a \hbox{0.4 arcsec} slit in the red-optical (R $\sim 93,000$), with a spectral range of \hbox{0.6447 \micron}\,-\,\hbox{1.0252 \micron}. \gjsn was observed on the first half of each night, while the second half was dedicated to observations of the GJ 65AB binary system. With the \hbox{2.18 arcsec}\ separation in 2015 September, \gjsfa and \gjsfb were observed simultaneously by placing both targets on the slit.
%With a separation of 2.18 arcsec\ in 2015 September both \gjsfa and \gjsfb could be observed simultaneously by placing both targets on the slit. 
Favourable seeing conditions throughout the two nights, with typically \hbox{$< 1$ arcsec}\ seeing, resulted in good spatial separation of the spectral profiles. During extraction, each observation was checked individually to ensure that the spatial extent of the two profiles were appropriately defined to avoid cross-contamination. The spectra were extracted using optimal extraction \citep{horne86extopt} with the Starlink package, {\sc echomop} \citep{mills14echomop}. A total of 86 and 77 spectra of \gjsn were obtained on each respective night using 180 sec exposures. For \gjsfa and \gjsfb, 91 and 87 spectra with 180 sec exposures were obtained.  

\begin{figure*}
   \centering
   \begin{tabular}{ccc}
      \hspace{4mm} \includegraphics[width=0.21\linewidth]{spec_gj791_25adj_div.ps} \hspace{4mm} & 
      \hspace{4mm} \includegraphics[width=0.21\linewidth]{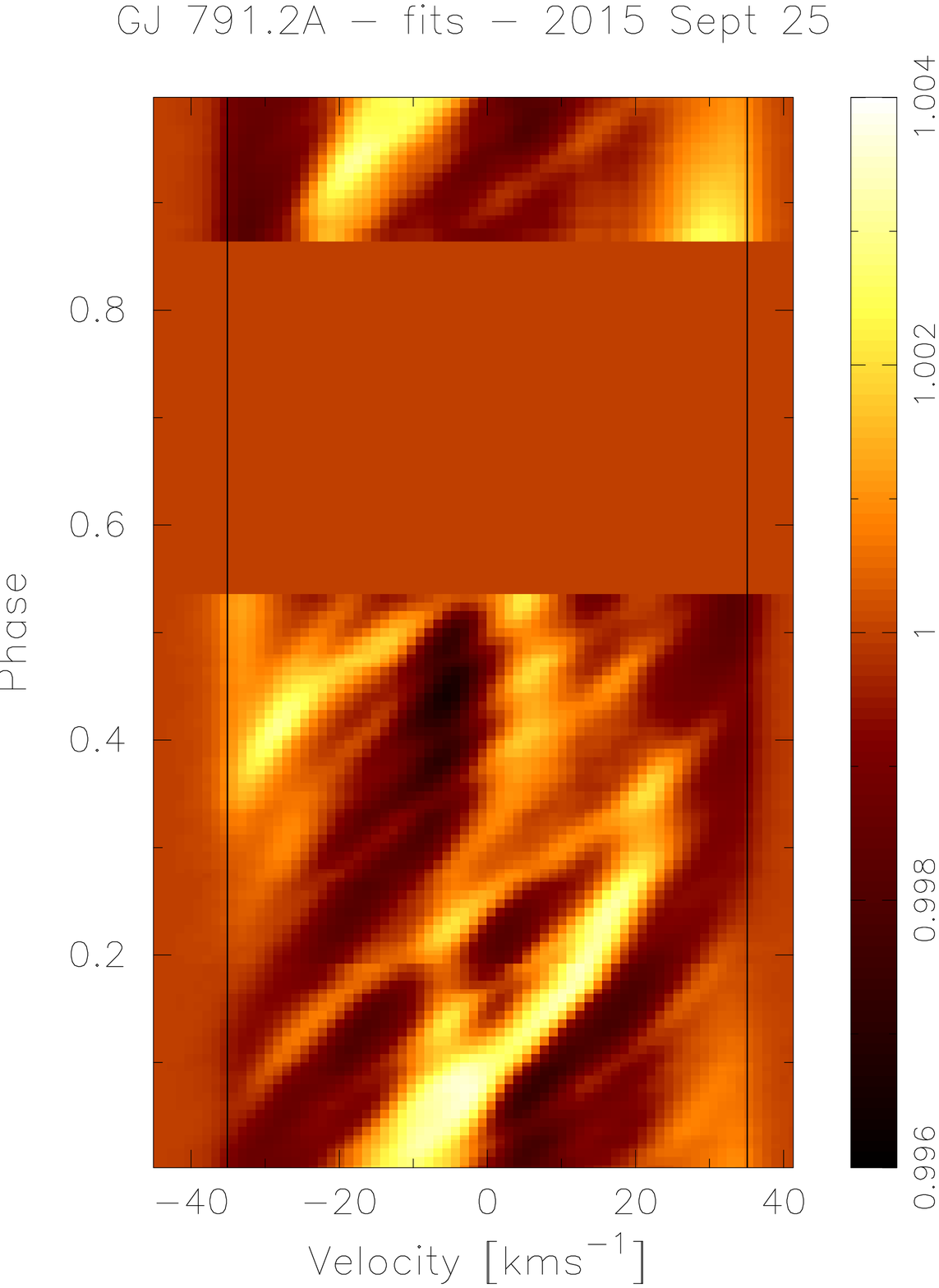} \hspace{4mm} &
      \hspace{4mm} \includegraphics[width=0.21\linewidth]{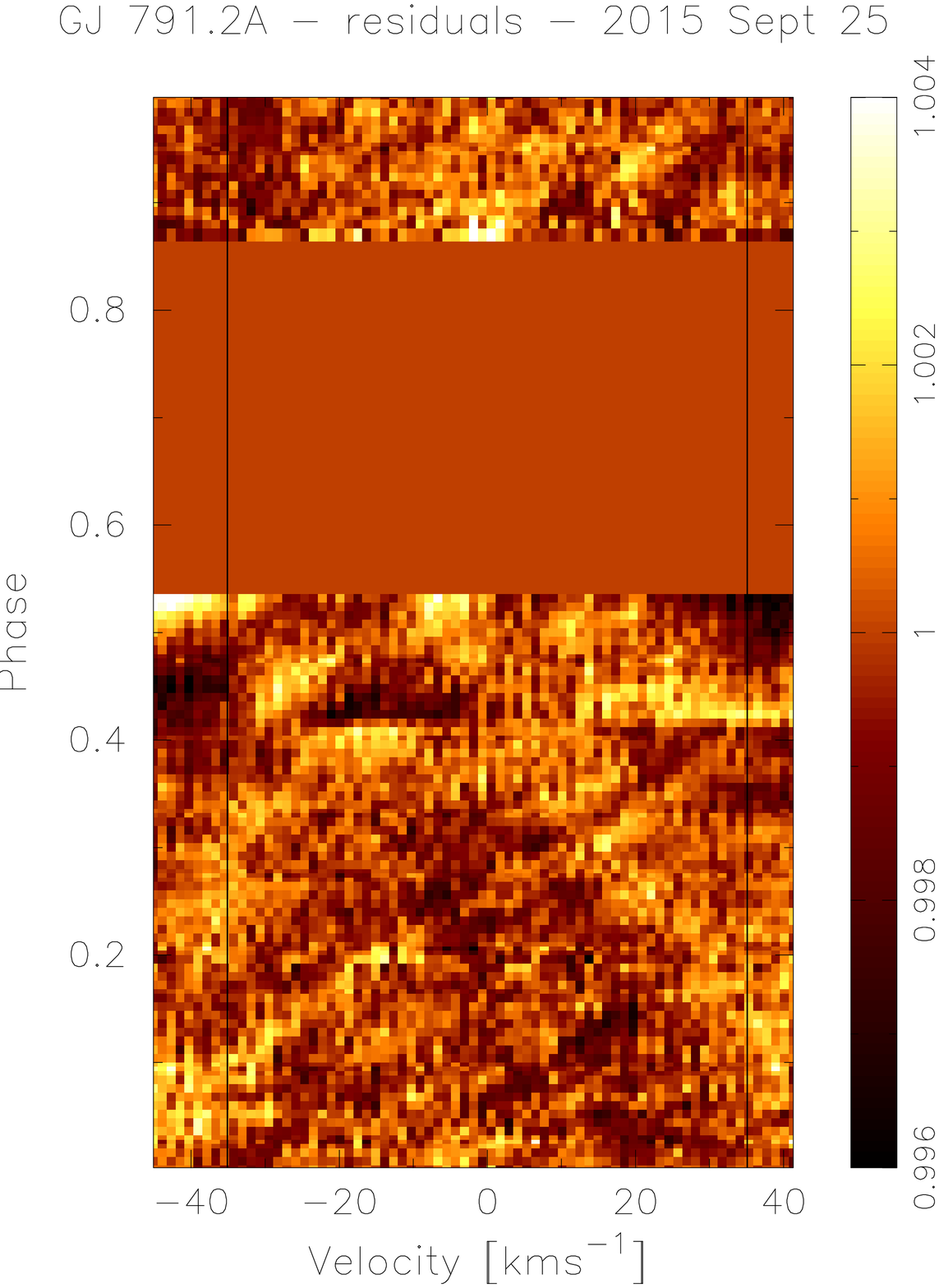} \hspace{4mm} \\
      \includegraphics[width=0.21\linewidth]{spec_gj791_28_div.ps} & 
      \includegraphics[width=0.21\linewidth]{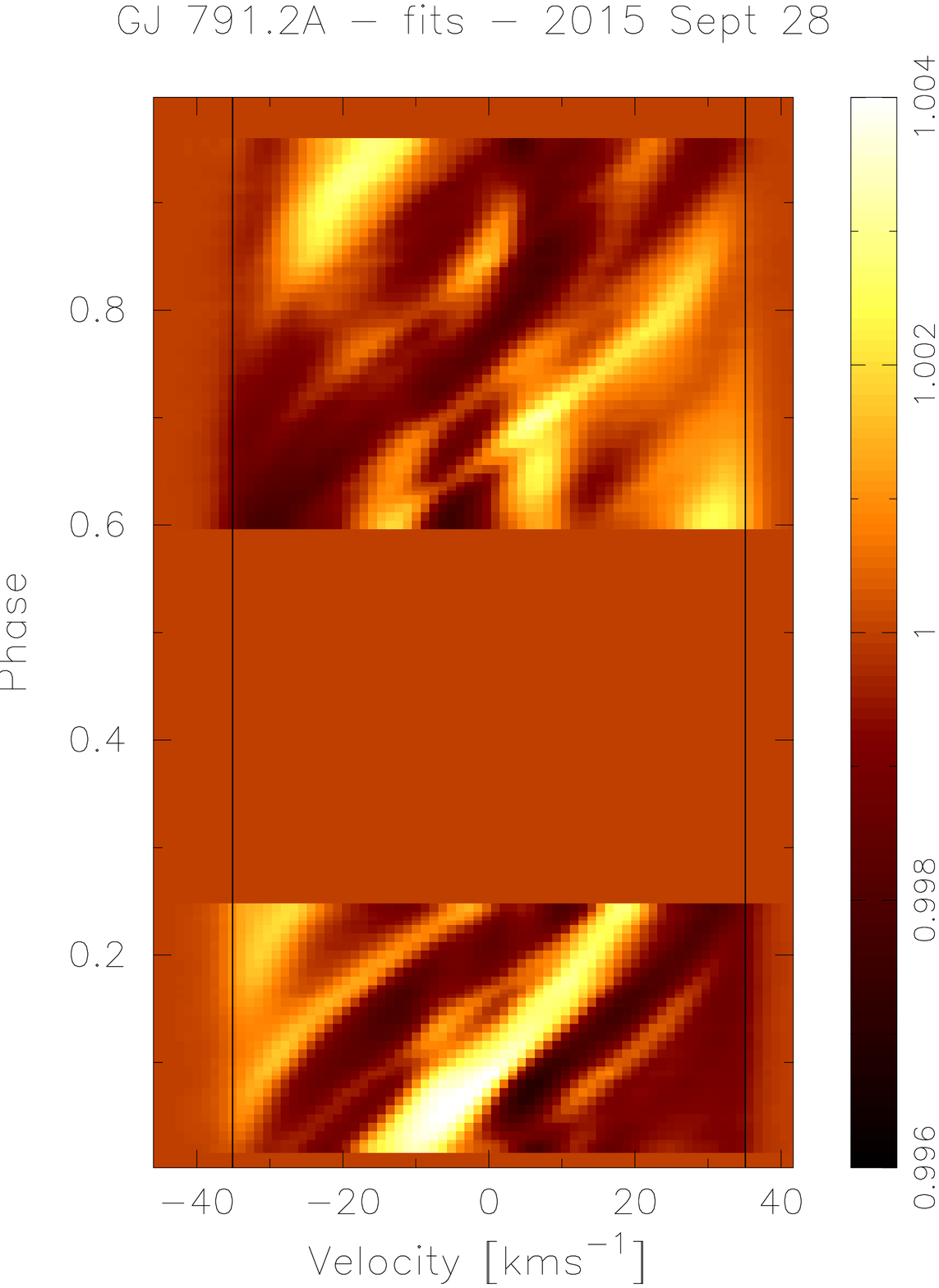} &
      \includegraphics[width=0.21\linewidth]{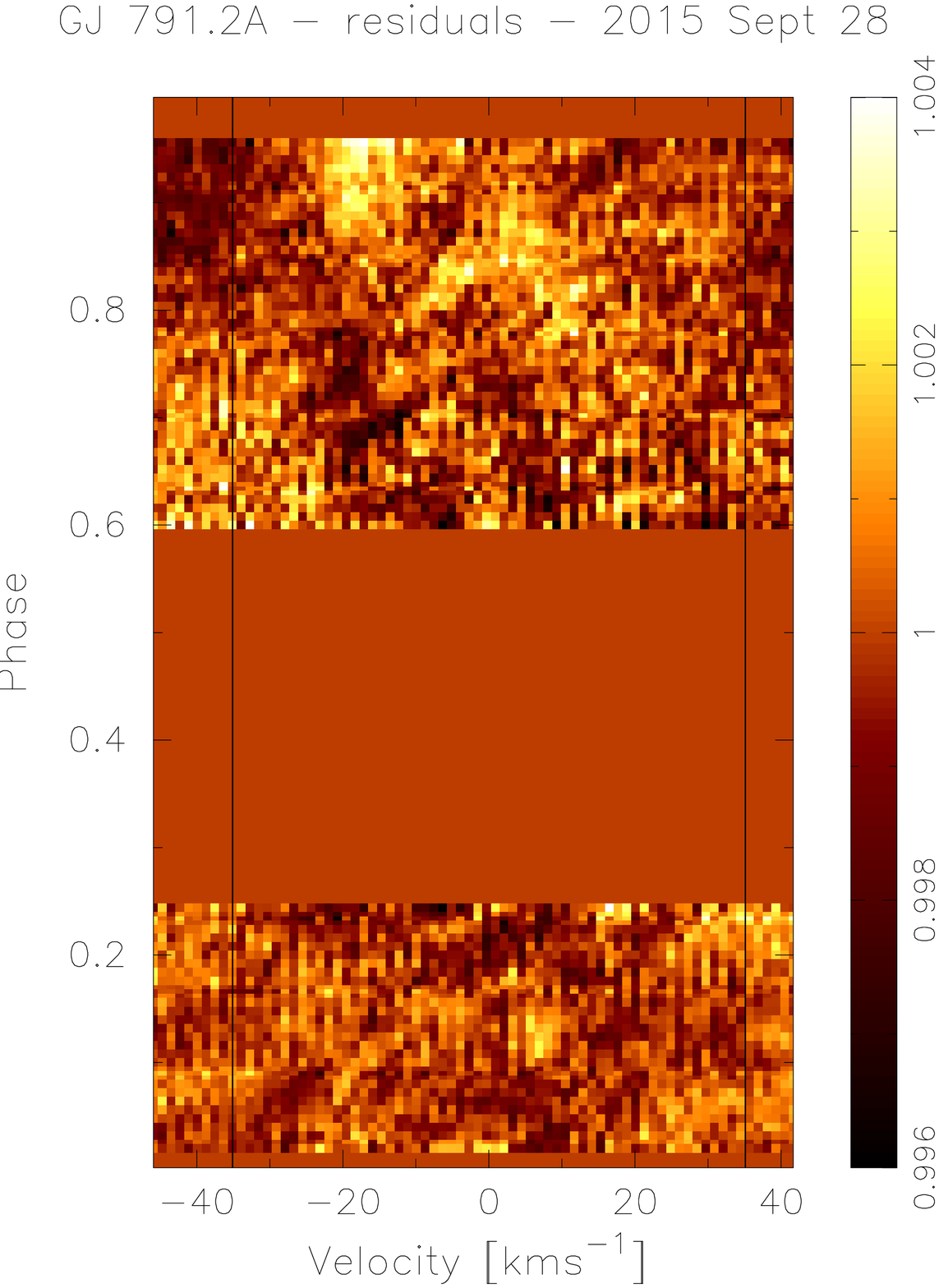} \\
   \end{tabular}
\caption{Phased residual spectral time series spectra of \gjsn (HU Del) for 2015 September 25 and 28. The deconvolved spectral profiles (left panels) have been divided by the mean profile; starspot trails appear white. The fits made to the spectra on September 25 and 28 individually (middle panels) and the corresponding residuals are shown (right panels). The fits correspond to image reconstructions in the upper two panels of Fig. \ref{fig:image_gj791}. The vertical lines denote the projected equatorial rotation velocity, \vsini\ = \hbox{35.1 \kms.}}
\label{fig:timeseries_gj791}
\end{figure*}

\begin{figure}
\begin{center}
\begin{tabular}{c}
\includegraphics[width=80mm,angle=0]{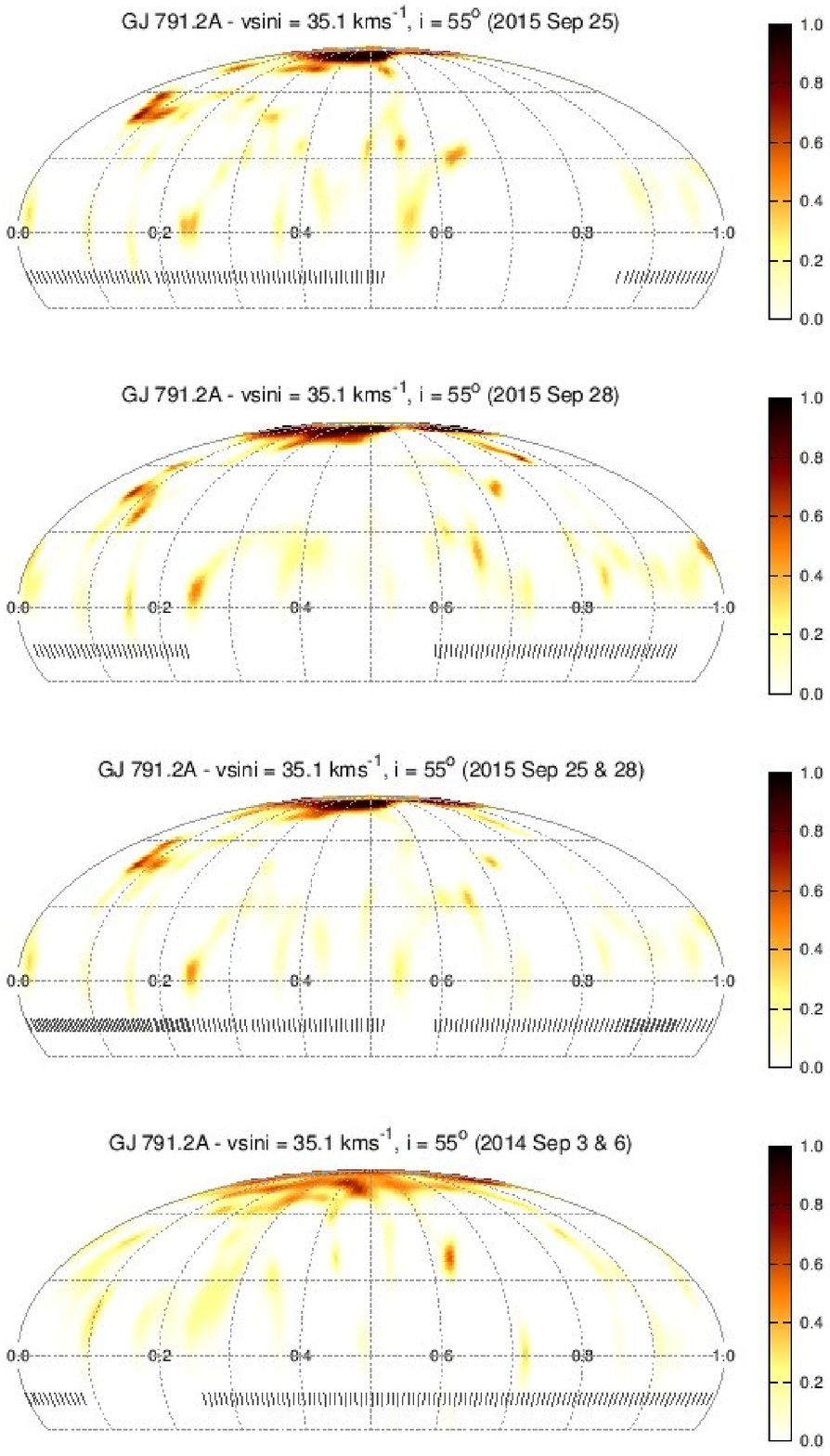} \\
\includegraphics[width=80mm,angle=0]{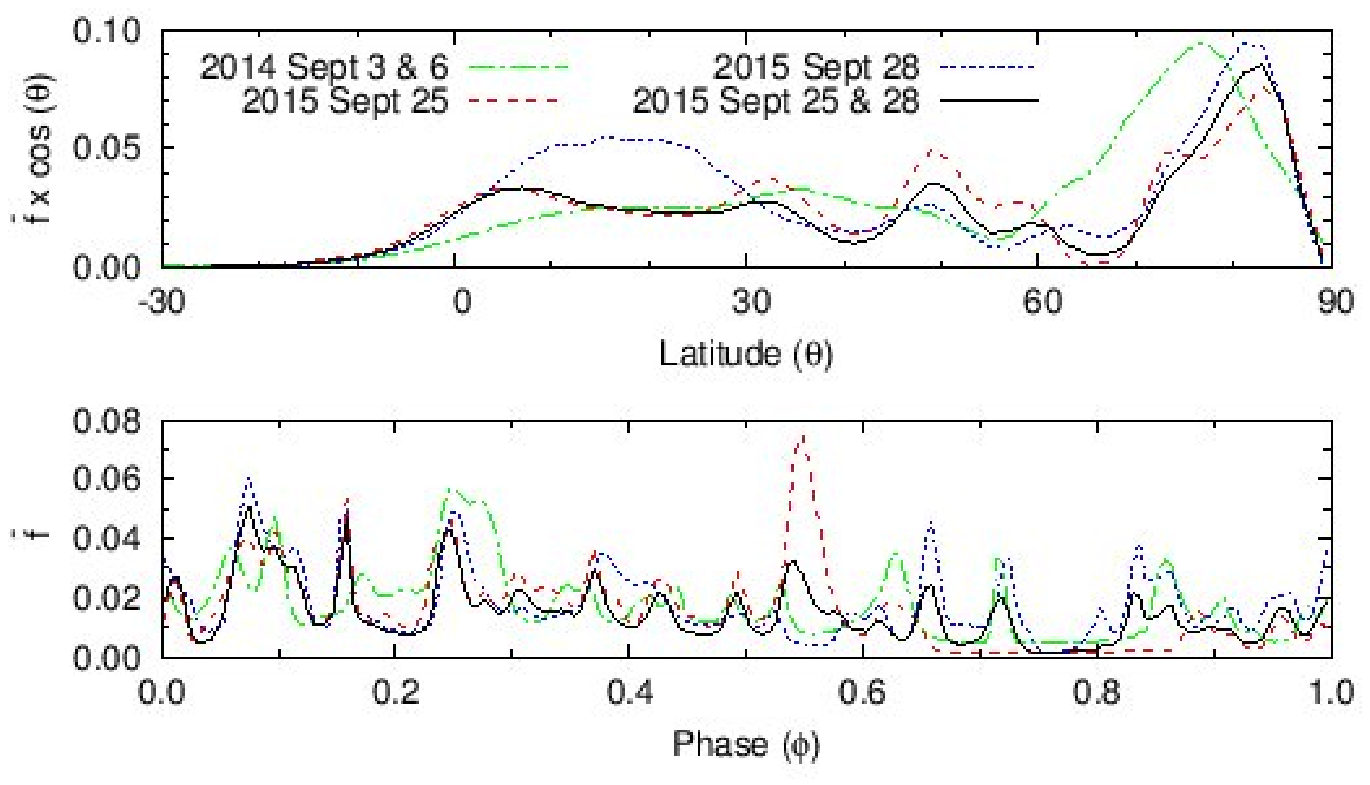} \\
%\hspace{-4mm}
%\includegraphics[width=45mm,angle=270]{rect_gj791_01mar_25_withstartimage_180scale.ps} \\
%\vspace{-13mm} \\
%\hspace{-4mm}
%\includegraphics[width=45mm,angle=270]{rect_gj791_01mar_28_withstartimage_180scale.ps} \\
%\vspace{-13mm} \\
%\hspace{-4mm}
%\includegraphics[width=45mm,angle=270]{rect_gj791_01mar_180scale.ps}  \\
%\vspace{-13mm} \\
%\hspace{-4mm}
%\includegraphics[width=45mm,angle=270]{rect_gj791_01mar_Reim2014_2600K_180scale.ps}  \\
%\vspace{-14mm} \\
%\hspace{-3mm}
%%\includegraphics[width=48mm,angle=270]{gj791_22jan_Reim2014_latlong.ps}  &
%\includegraphics[width=48mm,angle=270]{gj791_30jan_latlong_2014_2015_coslat.ps}  \\
\end{tabular}
\end{center}
\caption{Starspot maps of \gjsn for 2015 September 25 and 28, combined image, and 2014 September 3 and 6 image. The maps are made using Mollweide or equal-area projections. Rotation phase is indicated (longitude $0\degs < l \leq 360\degs$ runs in the opposite sense to phase, from right to left). Tick marks below the equator indicate the phases at which observations were made. Latitudes from $-30\degs \leq \theta \leq 90\degs$ are plotted. The images are phased to the midpoint of the first exposure, HJD0 = 2456904.636013, for consistency with the images in B15. The bottom panels show the mean latitude and mean phase spot filling (area corrected) for each map.}
\protect\label{fig:image_gj791}
\end{figure}

\subsection{Least squares deconvolution}
\protect\label{section:lsd}

Least squares deconvolution was applied to each spectrum to derive a single line profile from the several thousand absorption transitions. Our implementation of the procedure \citep{barnes98aper}, first described in \citet{donati97zdi}, was applied to M dwarf spectra using empirically derived line lists in \citet{barnes12rops}. The line lists were derived from observations of GJ 105B (M4.5V) (B15) and $4 \times 300$ sec observations of the slowly rotating star, GJ 1061 (M5.5V), made on 2015 September 25 and 28. In B15, we used the same procedure to derive time series spectra and perform Doppler imaging of \gjsn and the M9 dwarf LP 944-20. It is important to exclude lines that do not arise in the photosphere; specifically telluric bands and those lines with a strong chromospheric contribution, including \ha, \ionjb{He}{1}, the infrared \ionjb{Na}{1} lines and \ionjb{Ca}{2} triplet are removed before deconvolution is carried out. Any photospheric lines adjacent to these chromospheric lines that fall within the velocity range over which deconvolution is performed are also excluded. This procedure is particularly important in active M dwarfs where large chromospheric emission variability is seen during flaring events. At the start of observations, \gjsfb was undergoing a strong flaring event. The deconvolved line shapes were strongly distorted, becoming asymmetric with increased equivalent width, necessitating that the first 24 spectra be excluded in the subsequent imaging procedure. A weaker flare was also seen on \gjsn at the end of the first night and which can be seen as a continuum tilt in the deconvolved time series spectra. We retained the affected spectra and corrected the continuum tilt. Table \ref{tab:targets} summarises the observations: the input S/N ratios over the range used for deconvolution and the effective S/N of the mean deconvolved profiles are listed, indicating effective gains in S/N of 56\,-\,57 compared with a single line.

\subsection{Doppler imaging fully convective stars with a two-temperature model}
\protect\label{section:di}

{ As with stars with higher $T_{\rm eff}$, and the fully convective stars in B15, we have assumed that a two-temperature model can adequately describe the temperature inhomogeneities on active stars.} We applied the two-temperature, maximum-entropy regularised imaging algorithm, DoTS \citep{cameron01mapping} to recover Doppler images of our targets. DoTS uses a spot filling factor, $f_i$ (taking values in the range 0.0\,-\,1.0), for each image pixel, $i$. Since the absorption lines present in the spectra of mid-late M stars are dominated by molecular transitions, we investigated the behaviour of the line intensities and equivalent widths in B15. Synthetic spectra were computed using the BT-Settl model atmospheres of \citet{allard12} with the WITA code \citep{gadun97} which uses opacity sources listed in \citet{pavlenko07lp944}; see also \citet{pavlenko14} and \citet{pavlenko15}. To determine intensity ratios and centre-to-limb variations in the continua and line equivalent widths for the appropriate effective temperatures, the model spectra, calculated for different limb angles, were interpolated onto the observed wavelengths and multiplied by the blaze function. Deconvolution was then performed in exactly the same way as for the observed spectra. Since the synthetic spectra are not a perfect match to the observed spectra of each of our targets, we used appropriate line lists derived from the synthetic spectra. Here we applied the same procedure to \gjsfa and \gjsfb as that described more fully in B15. In addition, the local intensity profile used for Doppler imaging was derived from the same slowly rotating stars, \hbox{GJ 105B} and \hbox{GJ 1061}, from which we derived the empirical line lists.

%a slowly rotating star of the same or similar spectral type. For this purpose, we used the same slowly rotating stars, \hbox{GJ 105B} and \hbox{GJ 1061}, from which we derived the empirical line lists.
%(see \S \ref{section:observations} above).

%%%%%%%%%%%%%%%%%%%%%%%%%%%%%%%%%%%%%%%%%%%%%%%%%%%%%%%%%%%%%%%%%%%%%%%%%%%%%%% RESULTS %%%%%%%%%%%%%%%%%%%%%%%%%%%%%%%%%%%%%%%%%%%%%%%%%

\section{Results}
\protect\label{section:results}

We find optimal fitting parameters, including
%radial velocity, $v_r$, 
axial inclination, $i$, rotation period, $P_{\rm rot}$, equatorial rotation velocity, \vsini, and differential rotation shear, $\Delta\Omega$ (\S \ref{section:diffrot}) by minimising \chisq\ { using a fixed number of iterations \citep{barnes01aper}}.\ A summary of properties for each target derived from Doppler imaging are given in Table \ref{tab:properties}.

\begin{figure*}
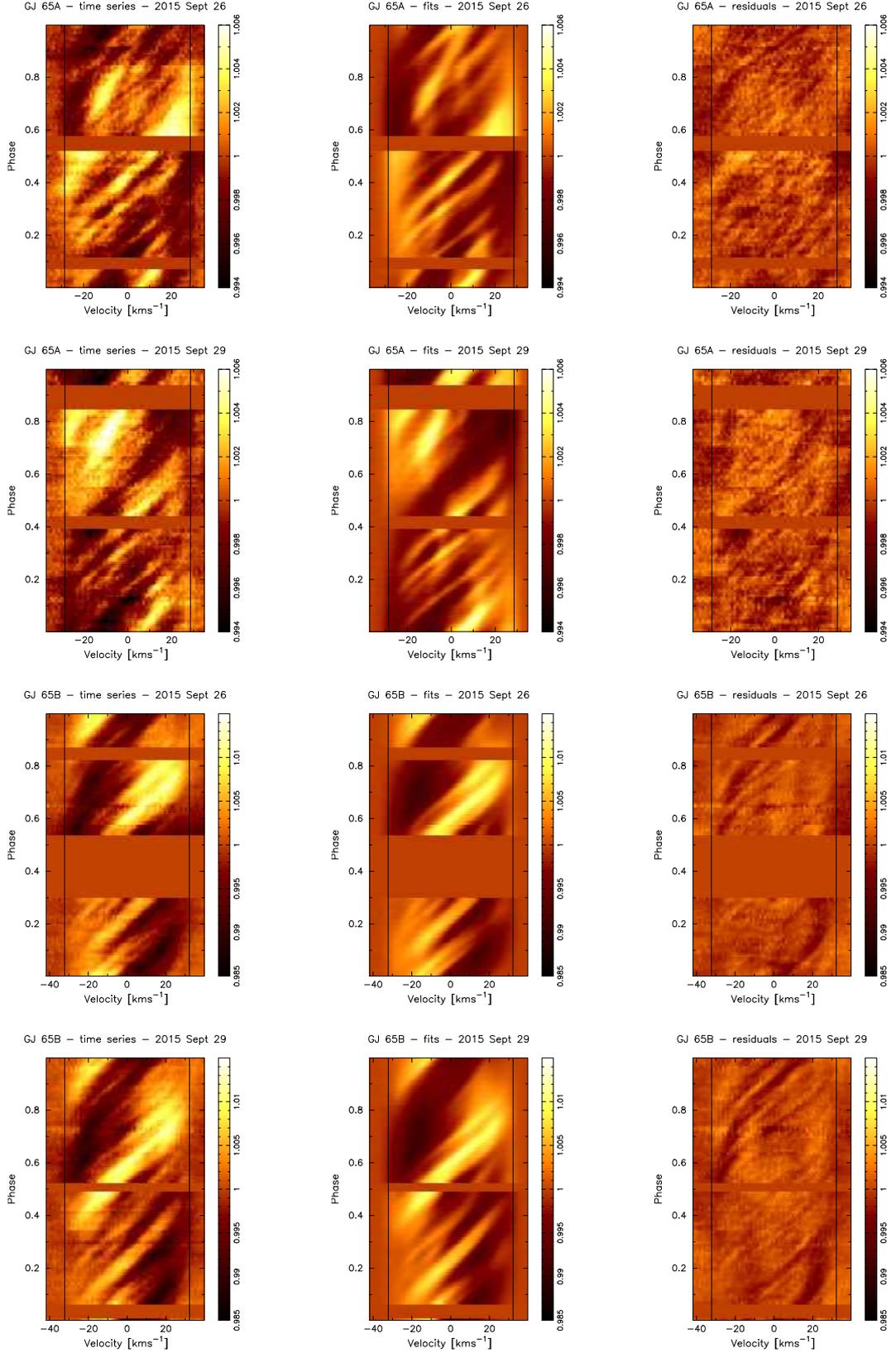

   \centering
   \begin{tabular}{ccc}
      \hspace{4mm} \includegraphics[width=0.21\linewidth]{spec_gj65a_25_div.ps} \hspace{4mm} & 
      \hspace{4mm} \includegraphics[width=0.21\linewidth]{fits_gj65a_30jan_25_div.ps} \hspace{4mm}  &
      \hspace{4mm} \includegraphics[width=0.21\linewidth]{res_gj65a_30jan_25_div.ps} \hspace{4mm} \\
      \includegraphics[width=0.21\linewidth]{spec_gj65a_28_div.ps} & 
      \includegraphics[width=0.21\linewidth]{fits_gj65a_30jan_28_div.ps} &
      \includegraphics[width=0.21\linewidth]{res_gj65a_30jan_28_div.ps} \\

      \includegraphics[width=0.21\linewidth]{spec_gj65b_25_div.ps} & 
      \includegraphics[width=0.21\linewidth]{fits_gj65b_30jan_25_div.ps} &
      \includegraphics[width=0.21\linewidth]{res_gj65b_30jan_25_div.ps} \\
      \includegraphics[width=0.21\linewidth]{spec_gj65b_28_div.ps} & 
      \includegraphics[width=0.21\linewidth]{fits_gj65b_30jan_28_div.ps} &
      \includegraphics[width=0.21\linewidth]{res_gj65b_30jan_28_div.ps} \\
   \end{tabular}
   \caption{Timeseries spectra of \gjsfa (rows 1 and 2) and \gjsfb (rows 3 and 4) as in Fig. \ref{fig:timeseries_gj791}. The respective projected equatorial rotation velocities, \hbox{\vsini\ = 28.6 \kms} and \hbox{32.2 \kms.} are indicated by the vertical lines. The time series are phased to the reference epoch of HJD0 = 2456556.9332.}
   \label{fig:timeseries_gj65}
\end{figure*}

\subsection{GJ 791.2A (HU Del) - 2015 September 25 and 28}
\protect\label{section:results_gj791_2015}

Our 2015 September 25 and 28 observations comprised better phase overlap than our prior 2014 September 3 and 6 observations, which were the basis of the work presented in B15. Since the 2014 observations were curtailed on the first night by weather constraints, we have re-derived the system parameters for \gjsn. In B15, phase overlap during the two nights of observations was limited to a narrow range of $\phi = 0.0000 - 0.0008$ from which we found \hbox{$P = 0.3088 \pm 0.0058$ d}, \hbox{\vsini = $35.1 \pm 0.4$ \kms} and \hbox{$i = 54$\degs\ $\pm\ 9$\degs}. With phase overlap of 0.336 for 2015 September 25 and 28, we find \hbox{$P = 0.3085 \pm 0.0005$ d} and \hbox{\vsini = $35.1 \pm 0.2$ \kms}. An axial inclination of $i = 55$\degs\ $\pm\ 4$\degs is found, indicating results in good agreement with our initial estimates. The phased time series spectra after subtraction of the mean profile are shown in Fig. \ref{fig:timeseries_gj791} (left panel) and reveal starspot features as white trails. Individual trails show different widths and gradients suggesting starspots or starspot groups of differing sizes at a range of stellar latitudes. The fits to the time series made separately for September 25 and September 28 are shown in the middle panels of Fig. \ref{fig:timeseries_gj791}. The residuals are plotted in the right hand panels and indicate the time series are fitted well. Some low level residual features remain and are still not fitted if we modify the full width of the local intensity profile for imaging. It is likely that imperfect fitting arises because the model does not account for bright chromospheric plage or intergranular faculae in the photosphere that may be associated with the starspots and which may additionally evolve on shorter timescales \citep{depontieu06}. A linear correction using the continuum regions was applied to the final 15 flare-affected observations on September 25, which caused the blue wing of the deconvolved profiles to appear depressed. 

%The fits necessitate use of a spot temperature of \hbox{$T_{\rm spot}$} \hbox{$= 2600$ K} to enable all spots to be modelled with spot filling such that $0.0 \leq f \leq 1.0$. 
The deviations from the mean line profile caused by the spots have sufficient amplitude that \hbox{$T_{\rm phot} - T_{\rm spot}$} \hbox{$= 400 K$} between the spotted and unspotted regions is required to fit the line profiles. This enables images with spot filling factors of $0.0 \leq f \leq 1.0$ to be recovered. For \hbox{$T_{\rm phot} =$} \hbox{$3000$ K} and \hbox{$T_{\rm phot} - T_{\rm spot}$} \hbox{$= 400$ K}, the continuum intensity contrast at disc centre at the mean wavelength of the deconvolved line profiles is \ispot/\iphot $= 0.32$. In the reconstructions for individual nights and the combined image in Fig. \ref{fig:image_gj791} (top and right panels), the greatest degree of spot filling is found in the circumpolar spot structure with maximum spot filling of  $f_{\rm max} = 0.98$ (98 per cent) on September 25. Spots with $\theta < 65$\degs typically possess spot filling factors of $0.10 < f < 0.72$, while for $\theta > 65$\degs spot filling factors are $0.21 < f < 0.98$. 
%The bottom right panel of Fig. \ref{fig:image_gj791} shows that the mean latitudinal filling at low and intermediate latitudes is an order of magnitude lower than at circumpolar latitudes. Under the assumption that spot size distributions at all latitudes are uniform, a ${\rm cos}(\theta)$ normalisation correction to the spot filling factors still indicates $\sim 3$ times more filling at higher latitudes. 
The bottom panels of Fig. \ref{fig:image_gj791} show that the mean latitudinal filling at low and intermediate latitudes is half that at circumpolar latitudes. The spot filling has been area-corrected by multiplying by cos($\theta$) so that a spot of fixed radius yields the same spot filling at all latitudes. Without the cos($\theta$) area correction, the spot filling is an order of magnitude higher in the circumpolar region.
The individual and combined reconstructions show that spots appear to be located at a range of latitudes and longitudes. The individual maps were reconstructed using the combined September 25 and 28 map as a starting image. This minimises the differences between images where there is no phase coverage and hence a weak constraint on the image. There appears to be some spot evolution where observation phases are common to both September 25 and September 28; for example the spot group centred on $\phi = 0.083$ and $\theta = 40\,-60$\degs has changed morphology. A mean spot filling factor of $\bar{f} = 0.023$ (2.3 per cent) is found for the combined image, while the 0.860 - 1.000 and 0.000 - 0.232 phase range common to the individual images yield mean respective spot filling factors for September 25 and September 28 of $\bar{f} = 0.023$ and $0.026$.

%\begin{figure*}
%\begin{center}
%\begin{tabular}{cc}
%
%\includegraphics[width=45mm,angle=270]{rect_gj65a_27jan_25_180scale.ps}  &
%\includegraphics[width=45mm,angle=270]{rect_gj65b_22may_25_180scale.ps}  \\
%%\includegraphics[width=45mm,angle=270]{rect_gj65b_20jan_25_180scale.ps}  \\
%\vspace{-13mm} \\
%\includegraphics[width=45mm,angle=270]{rect_gj65a_27jan_28_180scale.ps}  &
%\includegraphics[width=45mm,angle=270]{rect_gj65b_22may_25_180scale.ps}  \\
%%\includegraphics[width=45mm,angle=270]{rect_gj65b_20jan_25_180scale.ps}  \\
%\vspace{-13mm} \\
%\includegraphics[width=45mm,angle=270]{rect_gj65a_27jan_180scale.ps}  &
%\includegraphics[width=45mm,angle=270]{rect_gj65b_22may_180scale.ps}  \\
%%\includegraphics[width=45mm,angle=270]{rect_gj65b_20jan_180scale.ps}  \\
%\vspace{-14mm} \\
%\includegraphics[width=48mm,angle=270]{gj65a_latlong_27jan.ps}  &
%%\includegraphics[width=48mm,angle=270]{gj65b_latlong_22may.ps}  \\
%\includegraphics[width=48mm,angle=270]{gj65b_latlong.ps}  \\
%
%\end{tabular}
%\end{center}
%\caption{GJ 65A and GJ 65 B as in Fig. \ref{fig:image_gj791}}
%\protect\label{fig:image_gj65old}
%\end{figure*}

\subsection{GJ 791.2A - 2014 September 3 and 6 re-analysis}
\protect\label{section:results_gj791_2014}

Since the improved phase overlap during the 2015 observations affords a more reliable estimate of rotation period and system parameters, we have re-derived the image for 2014 September 3 and 6 using the new parameters. The image is shown in Fig. \ref{fig:image_gj791} (left panel) where we have used the same spot temperature as for the 2015 reconstruction to enable a direct comparison of the two sets of images. In B15, we found that the data required \hbox{$T_{\rm phot} - T_{\rm spot}$} \hbox{$= 300$ K} (\ispot/\iphot $= 0.41$) to enable the spot features to be fit. This implies weaker spots were present in 2014; adopting the larger \hbox{$400$ K} that we require for the 2015 image will thus result in smaller spot filling fractions. The new image in Fig. \ref{fig:image_gj791} is broadly the same as the B15 image reconstruction, since the system parameters are almost identical. As expected, the mean and maximum spot filling factors are lower, with $\bar{f} = 0.028$ and $f_{\rm max} = 0.71$ (c.f. $\bar{f} = 0.033$ and $f_{\rm max} = 0.76$ for the optimised image in Fig. 7 of B15 with $T_{\rm spot} = 2700$ K). The latitude distribution of spots at both epochs are very similar, particularly at low and intermediate latitudes, although the polar and circumpolar spot structures are stronger but less extensive in 2016.

%{GJ 791.2A} is an M4.5 dwarf with a previously estimated $v$\ sin\ $i$ = $32.1 \pm 1.7$\ kms$^{-1}$ \citep{delfosse98mdwarfs}. At 8.84\ pc it is a bright, nearby binary system with an astrometrically determined period of $1.4731 \pm 0.0008$ yrs, a maximum separation of $\sim$0.16 arcsec\ {and primary component mass of 0.286\,\msun\  \citep{benedict00gj791}. Our spectra do not show evidence for the secondary \hbox{0.126\,\msun} component, which is estimated to be $\Delta V = 3.23$ fainter than the primary component. \citet{montes01members} found this young disk system does not satisfy Hyades Supercluster membership. 

%Hence the separation of the stars varies with periastron and apastron separations of 2.1 AU and 8.9 AU respectively. Because the orbital separation is large compared with the stellar radii, tidal effects will be small but variable.

\begin{figure*}
\begin{center}
\begin{tabular}{cc}

%\includegraphics[width=45mm,angle=270]{rect_gj65a_01mar_25_180scale.ps}  &
%%\includegraphics[width=45mm,angle=270]{rect_gj65b_22may_25_180scale.ps}  \\
%\includegraphics[width=45mm,angle=270]{rect_gj65b_01mar_25_180scale.ps}  \\
%\vspace{-13mm} \\
%\includegraphics[width=45mm,angle=270]{rect_gj65a_01mar_28_180scale.ps}  &
%%\includegraphics[width=45mm,angle=270]{rect_gj65b_22may_25_180scale.ps}  \\
%\includegraphics[width=45mm,angle=270]{rect_gj65b_01mar_28_180scale.ps}  \\
%\vspace{-13mm} \\
%\includegraphics[width=45mm,angle=270]{rect_gj65a_01mar_180scale.ps}  &
%%\includegraphics[width=45mm,angle=270]{rect_gj65b_22may_180scale.ps}  \\
%\includegraphics[width=45mm,angle=270]{rect_gj65b_01mar_180scale.ps}  \\
%\vspace{-14mm} \\
%\includegraphics[width=48mm,angle=270]{gj65a_latlong_01mar_coslat.ps}  &
%%\includegraphics[width=48mm,angle=270]{gj65b_latlong_22may.ps}  \\
%\includegraphics[width=48mm,angle=270]{gj65b_latlong_01mar_coslat.ps}  \\

\includegraphics[width=170mm,angle=0]{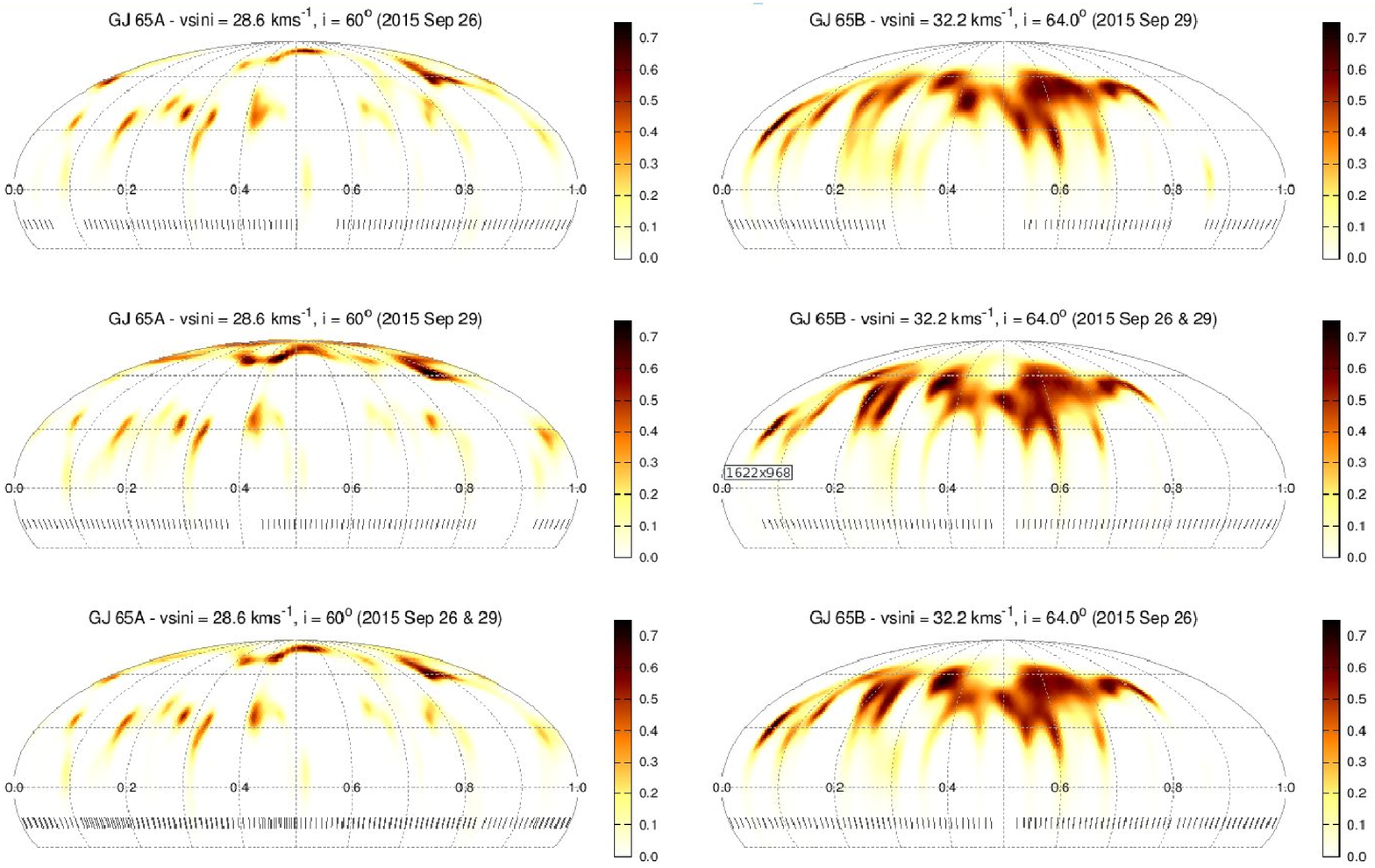}  \\
\includegraphics[width=170mm,angle=0]{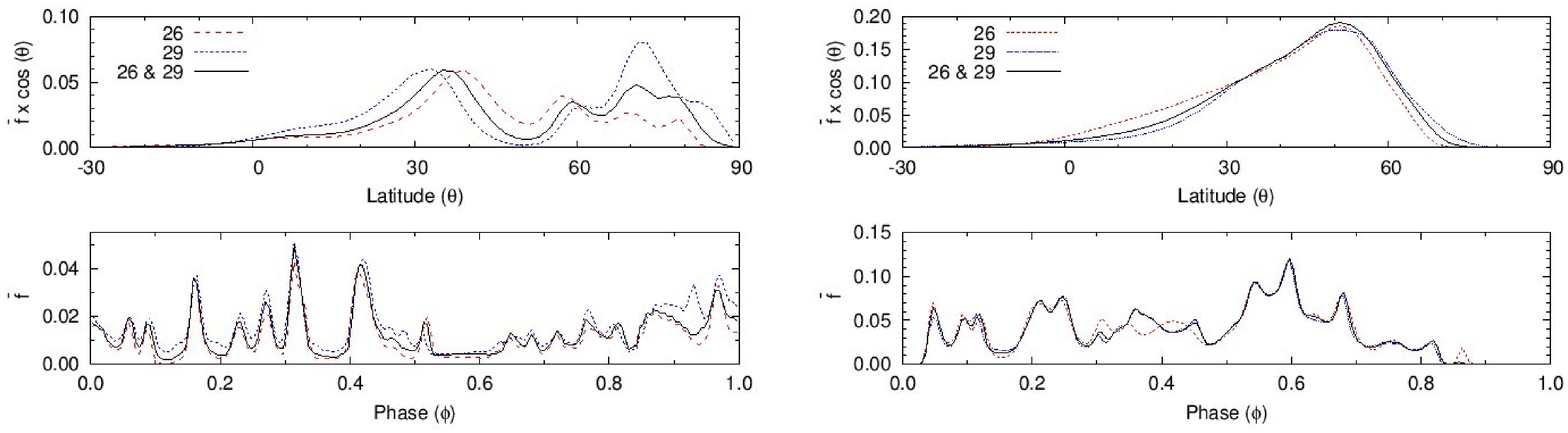}  \\

\end{tabular}
\end{center}
\caption{GJ 65A and GJ 65 B as in Fig. \ref{fig:image_gj791}}
\protect\label{fig:image_gj65}
\end{figure*}

\subsection{GJ 65A (BL Cet)}
\protect\label{section:results_gj65a}

We derived stellar images using synthetic spectra with \hbox{$T_{\rm phot} =$}\ \hbox{$2800$ K} and $T_{\rm spot} = 2400$ K (\ispot/\iphot $= 0.26$). This enables a direct comparison with \gjsfb, although we find that \hbox{$T_{\rm spot} = 2500$ K} (\ispot/\iphot $= 0.39$) also enables the spectra of \gjsfa to be fit with spot filling factors such that $f < 1.0$ in all image pixels. We find best fit parameters for \gjsfa of \hbox{\vsini =} \hbox{$28.6 \pm 0.2$ \kms}, $P = 0.2430 \pm 0.0005$ d (5.83 hrs; see \S \ref{section:diffrot}) 
%and $i = 59.2 \pm 1.2$\degs. 
and $i = 60 \pm 6$\degs. 
The mean profile subtracted time series spectra, fits and residuals are shown in the upper two rows of Fig. \ref{fig:timeseries_gj65}. We have used the same reference epoch (HJD0 = 2456556.9332) as \citep{kochulov17} who have recently published Zeeman Doppler images of \gjsfa and \gjsfb from polarimetric Stokes V observations made on 2013 September 21 and 24. We compare our maps with their findings in \S \ref{section:discussion}. The time series for the individual nights are similar, indicating that the same starspot features are persist on the 3 day timescale ($> 12$ stellar rotations) of the observations. There is nevertheless indication of evolution in some of the features. The time series model fits and residuals shown in \hbox{Fig. \ref{fig:timeseries_gj65}} 
%and Fig. \ref{fig:timeseries_profiles} 
indicate good fits though there are some residuals above the noise. For the individual and combined nights, 
% Left method 2 from make_snestimate gives normalisation factor of 2.11547. Values obtained via grep "chisq (spec)" dotsout_25.dat_30jan dotsout_28.dat_30jan dotsout_2528.dat_30jan | awk '{print $6/2.555**2}'
the respective reduced \hbox{\chisqr =} \hbox{$1.14,\ 1.38$ and $1.57$}. \hbox{Fig. \ref{fig:image_gj65}} (left panels) show the images for \gjsfa derived for the individual nights of September 26 and September 29 and also the image using data from both nights combined. Mean and maximum spot filling of $\bar{f} = 0.017$ and $f_{\rm max} = 0.65$ are found with spots located predominantly in a band centred at latitude $\theta \sim 30$\degs - $40$\degs. Higher latitude spots from $50\degs$ to $85$\degs are also found in the form of 2-3 larger spots or unresolved spot groups. There is some evidence of spot evolution between the two nights with a higher degree of polar spot filling on September 29, including additional spot structure. Otherwise the image reconstructions on September 26 and 29 are remarkably similar. Although high latitude spot structure is recovered no symmetric polar spot is seen.

\subsection{GJ 65B (UV Cet)}
\protect\label{section:results_gj65b}

The best fit parameters for \gjsfb are \hbox{\vsini =} \hbox{$32.2 \pm 0.2$ \kms}, $P = 0.2268 \pm 0.0003$ d (5.44 hrs) and $i = 64 \pm 7$\degs. As with \gjsfa, we used input models with $T_{\rm phot} = 2800$ K and $T_{\rm spot} = 2400$ K. The mean profile subtracted time series spectra, fits and residuals are shown in the lower two rows of Fig. \ref{fig:timeseries_gj65}. The starspot trails appear wider and more pronounced than those on \gjsfa (note colour scale), leading us to expect larger spots with greater contrast. The spot patterns are also consistent, with apparently little evolution between September 26 and 29. The residual fitted time series show that the features are well modelled, although again residuals are seen after subtracting the fits, likely due to associated facular contributions that we do not model. Mean and maximum spot filling of $\bar{f} = 0.056$ and $f_{\rm max} = 0.73$ are found.
% Left method 2 from make_snestimate gives normalisation factor of 2.555. Values obtained via grep "chisq (spec)" dotsout_25.dat_30jan dotsout_28.dat_30jan dotsout_2528.dat_30jan | awk '{print $6/2.555**2}'
For the individual and combined nights, we find \chisqr $ = 1.27,\ 1.20$ and $1.29$.
%Respective reduced \chisqr for individual and combined nights are \chisqr $ = 1.27,\ 1.20$ and $1.29$.
The images in Fig. \ref{fig:image_gj65} (right panels) reveal a contrast with \gjsfa, with particularly strong filling centred at mid-latitudes ($\theta \sim 50\,-\,56$\degs). There is a distinct lack of polar or circumpolar spot structure above $\theta \sim 70$\degs. Although large spot features appear at most phases, there is a notable lack of spots at
%$\theta \sim 0.55\,-\,0.77$\degs 
$0.00 \leq \theta \leq 0.04$ and $0.83 \leq \theta \leq 1.00$ (i.e. for 0.21 of the phase or 76\degs)
despite good phase coverage on both nights. The starspot pattern on \gjsfb is stable on the 3 night time scale of the observations, with little apparent evolution of spots after 13 stellar rotations.

%% DIFFERENTIAL ROTATION FOR EACH STAR - GJ791.2A, GJ65A and GJ65B
\begin{figure}
 \begin{center}
  \begin{tabular}{c}

    \includegraphics[width=55mm,angle=270]{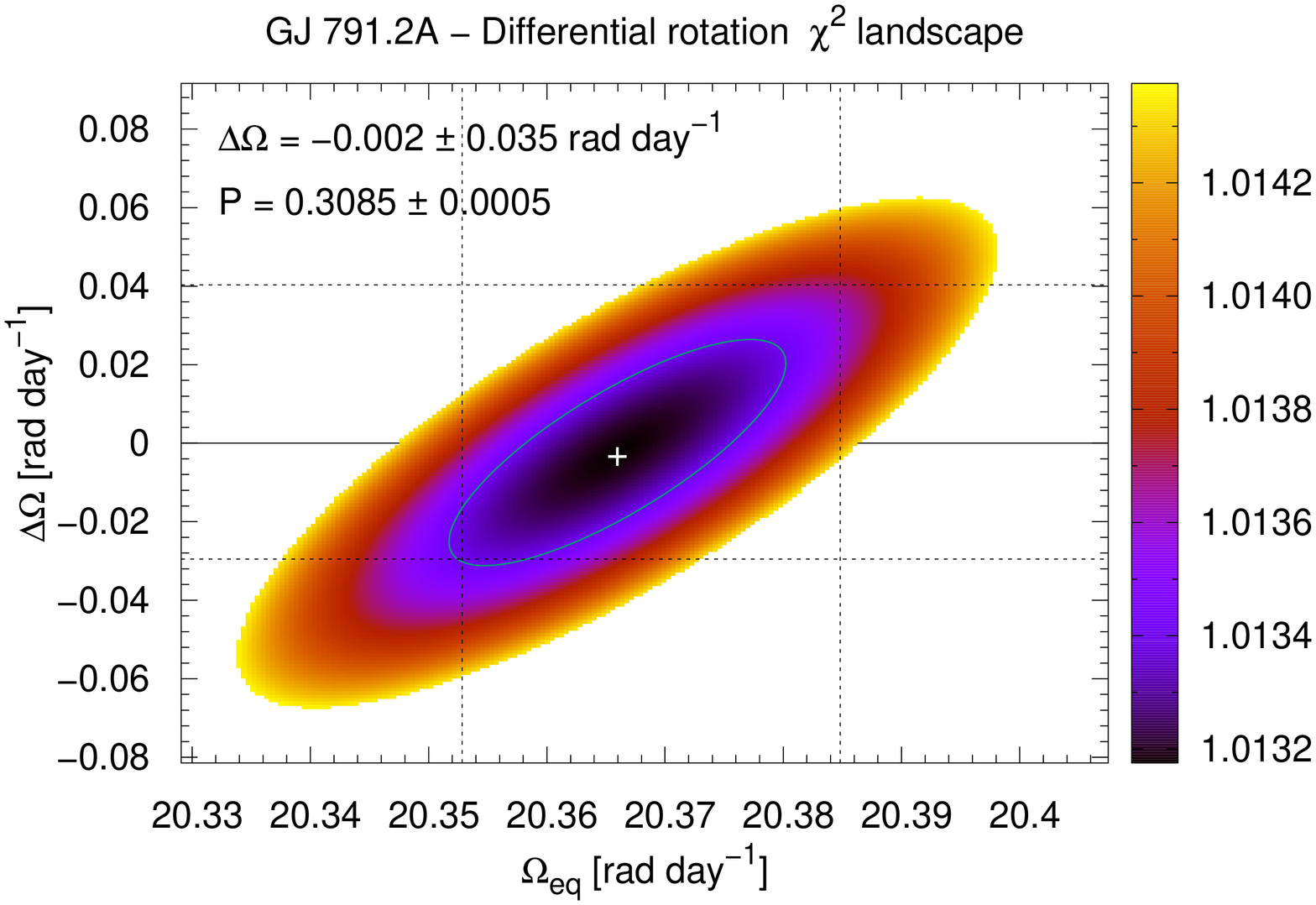}  \\
    \includegraphics[width=55mm,angle=270]{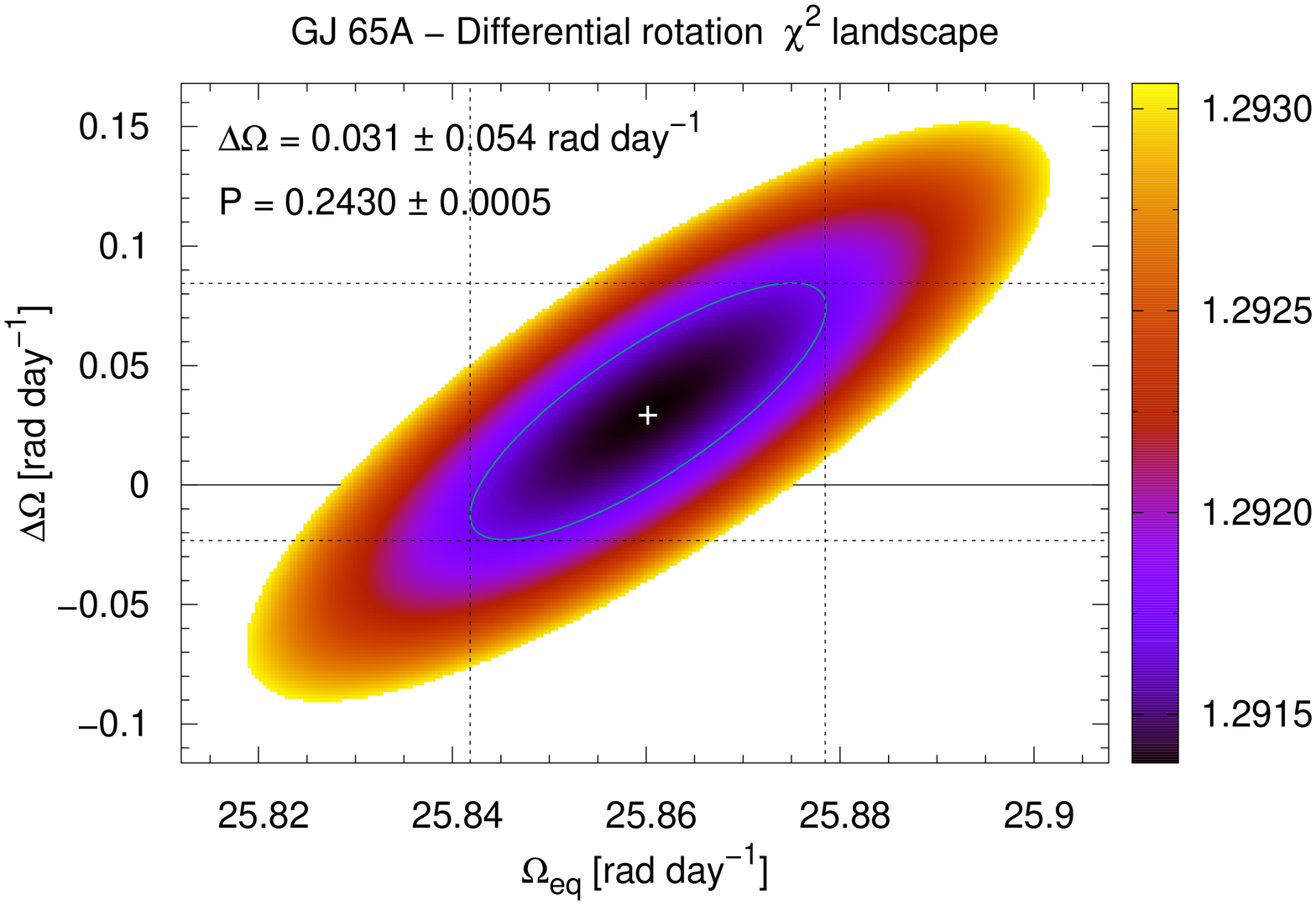}  \\
    \includegraphics[width=55mm,angle=270]{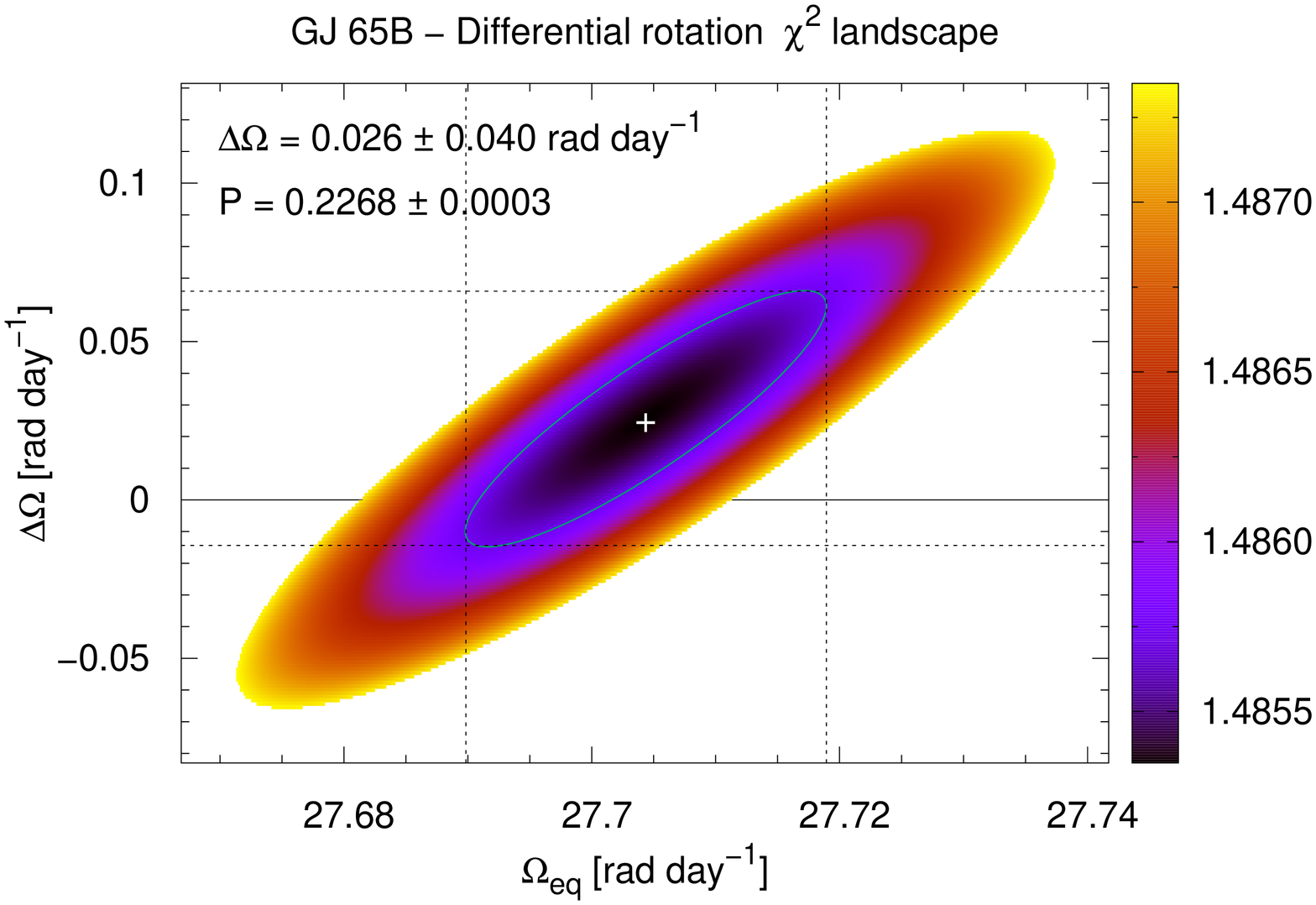} \\

  \end{tabular}
 \end{center}
 \caption{Estimation of differential rotation for \gjsn, \gjsfa and \gjsfb (top to bottom) assuming solar-like latitude dependent rotation. The \chisqr plots show shear, $\Delta\Omega$, plotted against equatorial angular rotation, $\Delta\Omega_{\rm eq}$, with the solid ellipse indicating the 2-parameter $1$-$\sigma$ contour. The dashed vertical and horizontal lines indicate the $1$-$\sigma$ uncertainties for each parameter while the solid line denotes $\Delta\Omega = 0$. Rotation period, $P_{\rm rot} = 2\pi/\Omega_{\rm eq}$ and $\Delta\Omega$ are indicated.}
\protect\label{fig:image_diffrot}
\end{figure}

\begin{figure*}
 \begin{center}
  \begin{tabular}{ccc}

    \includegraphics[width=49mm,angle=270]{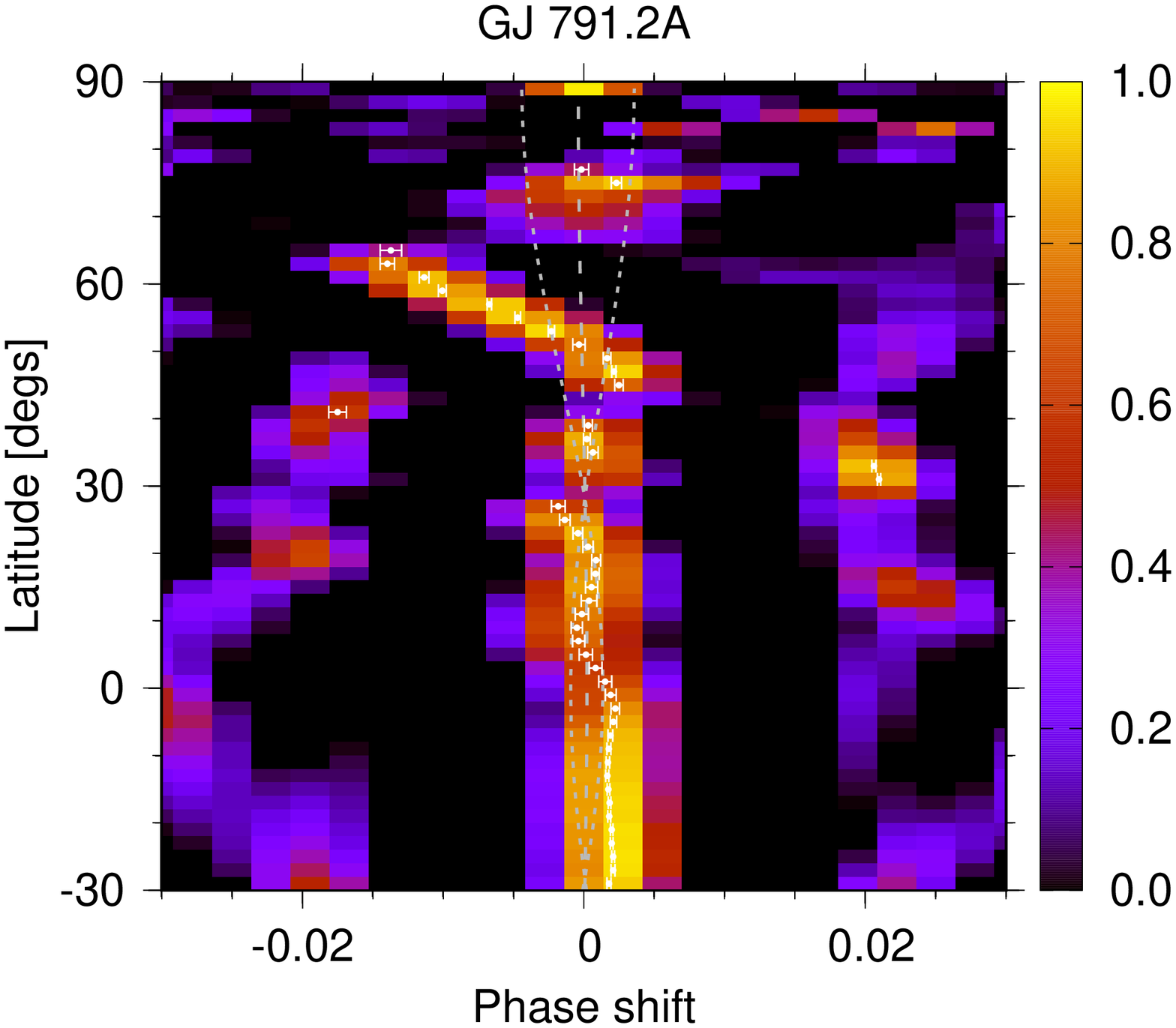}  & \hspace{-2mm}
    \includegraphics[width=49mm,angle=270]{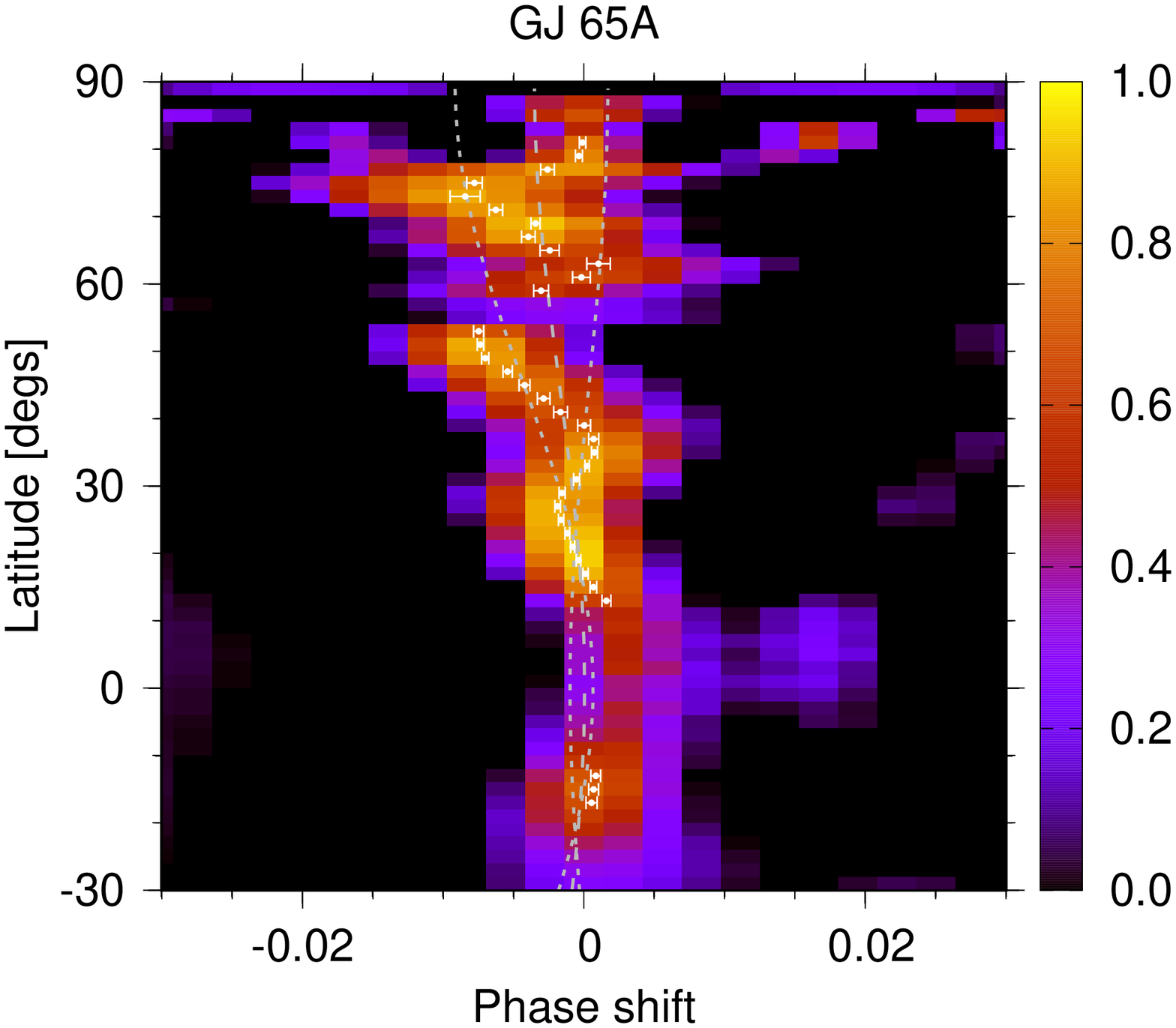}  & \hspace{-2mm}
    \includegraphics[width=49mm,angle=270]{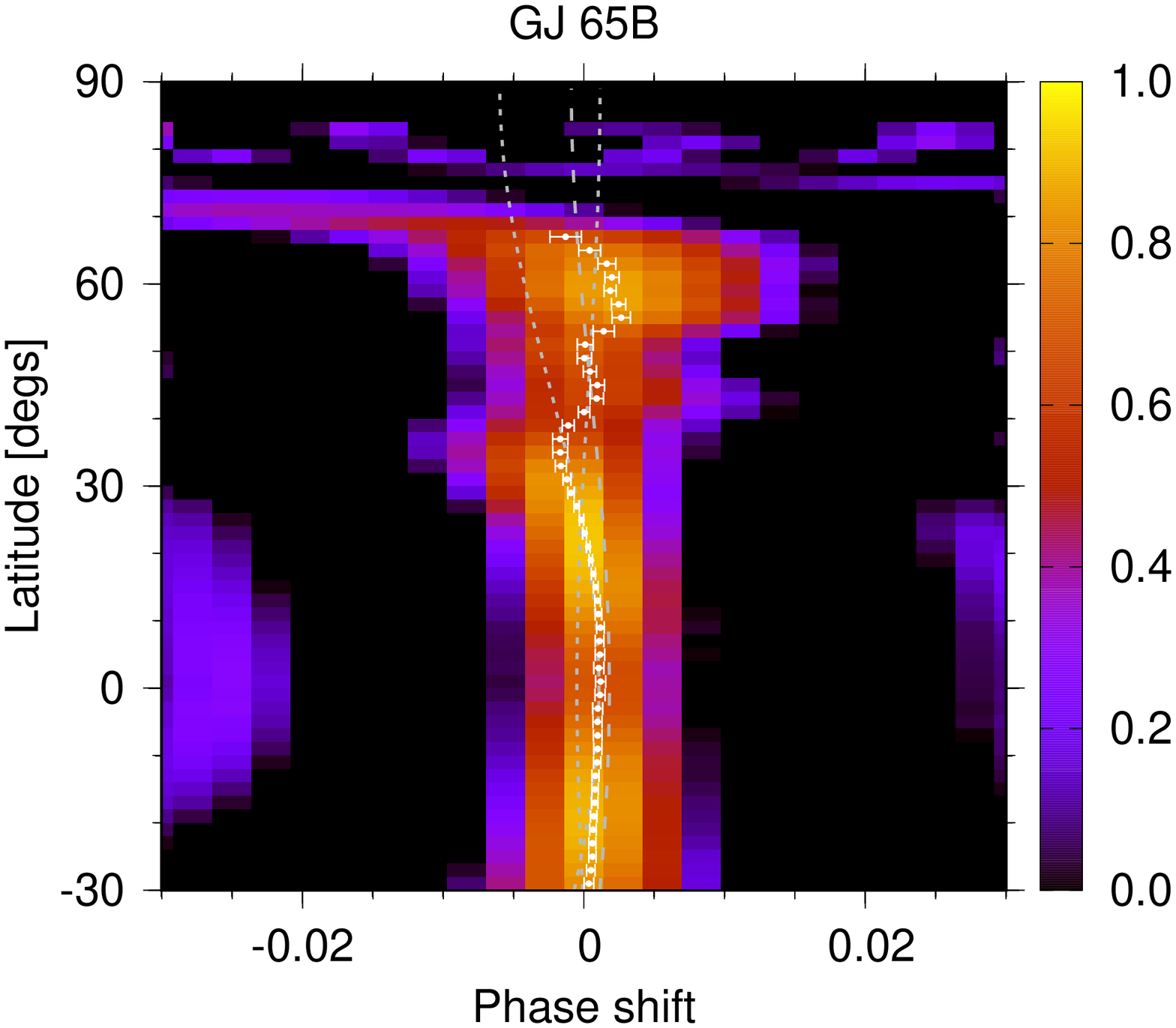} \\

  \end{tabular}
 \end{center}
 \caption{{ Latitudinal cross-correlation maps using September 25 and September 28 images for \gjsn, \gjsfa and \gjsfb. The white points and error bars show the maximum of the cross-correlation peak and uncertainty at each latitude. The dashed and dotted curves are the solar-like differential rotation and respective uncertainties obtained using the sheared image method parameters in Fig. \ref{fig:image_diffrot} for each target.}}
\protect\label{fig:image_latcross}
\end{figure*}

\subsection{Differential rotation estimation}
\protect\label{section:diffrot}

\subsubsection{{ Solar-like differential rotation via the sheared image method}}
\protect\label{section:diffrot_sheared}

The persistence of spot features in the time series spectra and Doppler images enables us to estimate the { latitude dependent} rotation using the sheared image method \citep{petit02}. We model the star as a differentially rotating body with a simple solar-like latitude dependence for the rotation

%differential rotation using the simple solar-like latitude dependence
%, with a solar-like sin$^2(\theta)$ law, where

\begin{equation}
\Omega(\theta) = \Omega_{\rm eq} - \Delta\Omega\,{\rm sin}^{2}(\theta)
\protect\label{eqn:dr}
\end{equation}

\noindent
where $\theta$ is the stellar latitude, $\Omega_{\rm eq}$ is the equatorial rotation rate ($P_{\rm rot} =$ $2\pi/\Omega_{\rm eq}$), and $\Delta\Omega = \Omega_{\rm eq} - \Omega_{\rm pole}$ is the magnitude of the shear.
%This enables recovery of $\Delta\Omega = \Omega_{\rm eq} - \Omega_{\rm pole}$. 
 The results of fitting for latitude dependent rotation are shown for each star in \hbox{Fig. \ref{fig:image_diffrot}}, where \chisqr\ is plotted for $\Delta\Omega$ vs $\Omega_{\rm eq}$. All three stars reveal small positive differential rotation, which is consistent with solid body rotation within the $1-\sigma$ uncertainties. The results are discussed further in \S \ref{section:discussion} in the context of other measurements and theoretical predictions.

\subsubsection{{ Departure from solar-like differential rotation?}}
\protect\label{section:diffrot_lat}

{
A recent publication by \cite{brun17diffrot} investigated the large scale flows in the convective envelopes of late-type stars as a function of rotation rate. Stars with small Rosby numbers (e.g. \hbox{$0.5$\,\msun}\ with rotation $> 3 \Omega_\odot$) show cylindrical angular rotation velocity profiles throughout the convection zone resulting in alternating zonal jets, where prograde and retorgrade flows are seen at the surface. The models of Brun et al. are for stars with a radiative core and convective envelope, and do not investigate the rotation rates of our targets which are two orders of magnitude greater than solar. 

To search for any departure from solar-like differential rotation for \gjsn, \gjsfa and \gjsfb, we performed cross-correlation of each latitude using the September 25 and September 28 Doppler images. This procedure is similar to the method first used to identify differential rotation in the K dwarf AB Dor \citep{donati97zdi} and Cluster G dwarfs \citep{barnes98aper}. Cross-correlation enables evolution of features and those features that trace any latitude dependent rotation to be assessed. We used a modified version of the {\sc hcross} algorithm of \cite{heavens93redshifts} to perform the cross-correlation (see \citealt{barnes12rops}). {\sc hcross} is part of the STARLINK package, {\sc figaro} \citep{shortridge93figaro}; now maintained by East Asian Observatory\footnote{http://starlink.eao.hawaii.edu}.
The results are shown in Fig. \ref{fig:image_latcross}. Globally, there appears to be reasonable agreement between the sheared image (assumed solar-like differential rotation) and the cross-correlation methods. For \gjsn, the overlapping phase range used for the cross-correlation was $\phi = 0.0\,-0.3$ and $0.8\,-\,1.0$, while the complete phase range was used for \gjsfa and \gjsfb. The white filled circles and horizontal bars denote the cross-correlation maxima and estimated uncertainties. The grey dashed curves show the differential rotation derived using Equation \ref{eqn:dr} via the sheared image method. The dotted grey curves show uncertainties in the sheared image differential rotation values. Cross-correlation of individual latitudes appears to show significant deviation from the solar-like differential rotation, but is not consistent between all three stars. For \gjsn, the phase coverage is smaller, with fewer features to obtain reliable cross-correlation. The apparent strong shearing at $\theta = 55$\degs, appears to be due to evolution of the spot group at phase $\phi = 0.1$ (see Fig. \ref{fig:image_gj791}), rather than rotational shearing of persistent spots. For \gjsfa, the lack of strong features at low latitudes results in cross-correlation peaks at larger phase shifts ($\phi \sim 0.1$). Hence the peak cross-correlations and uncertainties are not shown for latitudes $-11$\degs\ $\leq \theta \leq 11$\degs. Correlated deviations over several latitudes in the cross-correlation peaks for \gjsfa and \gjsfb are of amplitude $\Delta\phi \lesssim 0.01$ (3.6\degs). Despite the relatively small formal errors from {\sc hcross}, these deviations are somewhat smaller than the equatorial resolutions of $\sim 8$\degs to $10$\degs\ reported in \S \ref{section:discussion}. Without repeated observations it is not possible to discern whether shearing due to persistent zonal flows is present or whether evolution of spot features masks such an effect. 

Under the assumption that a differential rotation law can be incorporated into the model to which we fit the data, the magnitude of the shear should be more reliable than simple cross-correlation of constant latitude bands (see below) since all observations (i.e. on both nights in this instance) are modelled simultaneously. Although we can't rule out deviation from a solar-like law, for consistency with previous studies, we restrict further discussion in \S \ref{section:discussion} to the solar-like differential rotation measured using the sheared image method.}

%%%%%%%%%%%%%%%%%%%%%%%%%%%%%%%%%%%%%%%%%%%%%%%%%%%%%%%%%%%%%%%%%%%%%%%%%%%%%%% DISCUSSION AND SUMMARY 
%%%%%%%%%%%%%%%%%%%%%%%%%%%%%%%%%%%%%%%%%%%%%%%%%

\section{Discussion}
\protect\label{section:discussion}

We have reconstructed the starspot distributions on three fully convective stars: GJ 65A (M5.5V), GJ 65 B (M6V) and GJ 791.2A (M 4.5V). These observations were secured with the {\sc vlt} and have an unprecedented equatorial resolution of $\lesssim 10$\degs, bringing the total number of fully convective stars with surface brightness distributions to six.  The surprising result is that the two components of the binary system, \hbox{GJ 65}, have very different starspot distributions, despite having near-equal rotation and near-equal mass. \gjsfa shows high latitude and circumpolar spot coverage, while in contrast, \gjsfb shows large spots at intermediate latitudes.  The third star, GJ 791.2A, shows similar spot coverage to an image made a year earlier (B15) and with a spot distribution similar to that reconstructed for GJ 65A.

We are able to achieve unprecedented resolutions of 8.3\degs, 10.1\degs and 9.0\degs at the equator of the three fully convective stars, \gjsn, \gjsfa and \gjsfb, for the first time. The Doppler images reveal that numerous spots are distributed across their surfaces, exhibiting similarities, with spots located at low-intermediate latitudes. \gjsfb shows spots or spot groups confined only to mid-latitudes with a higher degree of spot filling compared with \gjsfa and larger spot sizes compared with both \gjsn and \gjsfa. 

\subsection{Spot filling}
\protect\label{section:discussion_spotfill}

We have investigated the issue of spot filling on \gjsfb by modifying the spot temperature. Spot filling is expected to increase when using a two-temperature model with smaller contrasts (i.e. larger values of \ispot/\iphot). For low contrasts, the spot areas may also increase as DoTS attempts to fit the line distortions with lower contrast spots by increasing their size. With a higher contrast of $T_{\rm phot} - T_{\rm spot}$ = \hbox{500 K} (\ispot/\iphot \hbox{$= 0.17$}), we find that the \chisqr is only marginally improved, by 3 per cent, over our adopted $T_{\rm phot} - T_{\rm spot}$ = \hbox{400 K} (\ispot/\iphot \hbox{$= 0.26$}). The images of \gjsfb look essentially the same for \hbox{$T_{\rm spot} = 2300$ K} and \hbox{$2400$ K}. Similarly, with an arbitrary setting of \ispot/\iphot \hbox{$= 0.1$}, the images are not changed significantly in appearance. Hence the difference in spot area sizes between \gjsfa and \gjsfb appears to be real and not a consequence of inappropriate choice of \ispot/\iphot. \gjsn closely resembles \gjsfa, but exhibits a greater degree of circumpolar spot structure. The structure appears to be slightly less complex in the 2015 images compared with the 2014 image where all spots show more uniform filling factors.

\subsection{System parameters and radii}
\protect\label{section:discussion_system}

Using our estimate of the axial inclination and period of \gjsfa and \gjsfb implies respective radii of \hbox{$R_A =$}\ \hbox{$0.159 \pm 0.010$ \rsun} and \hbox{$R_B =$}\ \hbox{$0.160 \pm 0.008$ \rsun}. The estimates are in good agreement with the \hbox{$R_A =$}\ \hbox{$0.165 \pm 0.006$ \rsun} and \hbox{$R_B =$}\ \hbox{$0.159 \pm 0.006$ \rsun} estimates by \citet{kervella16gj65} from astrometric and parallax measurements. $P_{\rm rot}$ and \vsini, are generally well constrained parameters, but our estimates of $i$, which contribute the dominant source of uncertainty in radius measurements (hence the likely reason for a larger radius estimate for \gjsfb), seem to be robust. \citet{kervella16gj65} suggest that the large radii are consistent with youth and further inflated by strong magnetic fields \citep{chabrier07,feiden12}. We discussed the radius and probable few hundred Myr age of \gjsn in B15. Our system parameters for \gjsn are in agreement with the previous estimates in B15 and now imply \hbox{$R = 0.261 \pm 0.013$}. 

\subsection{Comparison with models and magnetic field observations}
\protect\label{section:discussion_comparison}

Although we only recover surface brightness images, with no information about magnetic field strength, the strongest magnetic fields on stars with convective envelopes are expected in the recovered spots. It is thus reasonable to expect that spots act as tracers of dynamo activity, and to ask whether different spot patterns can be explained by different dynamo modes, or cyclic behaviour. Simulations by \citet{gastine13} find a dynamo bistability occurring in more rapidly rotating fully convective stars (i.e. with low Rossby numbers). The two dynamo modes suggest either dipolar or multipolar field configurations while \citet{yadav15} find that the axisymmetric dipolar mode is stable and dominant. The simulations by \citet{yadav15} also show heat flux maps, which offer the opportunity for comparison with our observations. Their simulations reveal stronger spots at high latitudes in the axisymmetric dipolar mode and are qualitatively in agreement with our images of \gjsfa, \gjsn and potentially the image of LP 944-20 (M9V) presented in B15. 

Our finding of higher spot filling factors and larger spots on \gjsfb compared with \gjsfa appears consistent with the stronger mean field strength found on \gjsfb by \citet{kochulov17}. However their observations suggest a strongly dipolar field for \gjsfb and more complex field for \gjsfa. The contrasting spot patterns may represent different parts of a magnetic cycle \citep{kitchatinov14}, although \citet{shulyak15dynamo} note that oscillating modes may only be identified by monitoring of individual M dwarfs on time scales of $\geq 15$ years. The simulated stability of the axisymmetric dipolar mode and the observation of consistently different radio behaviour in the \hbox{GJ 65} components noted by \citet{kochulov17} suggest that any cyclical behaviour must occur on timescales greater than 30 years.

%For \gjsfb, the radial field found by \citet{kochulov17} for \gjsfb appears to be strongest in the $0.25 \leq \phi \leq 0.50$ region and the $0.7 \leq \phi \leq 0.8$ (their Figure 1b) while the spots we reconstruct also span $0.05 \leq \phi \leq 0.82$ (Fig. \ref{fig:image_gj65} bottom right panel above). There may thus be some suggestion that the strongest magnetic regions are persistent on the two year timescale between our observations and the Stokes V observations, indicating long term stability of the dynamo mode. The lack of exact correspondence implies evolution of the individual spot features is likely to have taken place. However, we caution that comparing brightness images and magnetic maps is not straightforward owing to observational biases: the strongest spots may also be too dark to enable the polarisation to be reliably measured in their vicinity.

There is some indication of phase alignment between our brightness images and the Stokes V radial field images derived from observations two years earlier by \citet{kochulov17}. While this may tentatively suggest long term stability of the dynamo, it is not clear that the spot geometry and magnetic field reconstructions offer a fully consistent picture. Specifically, the brightness images of \gjsfb show only spots at mid latitudes. No spot structure at high latitudes or near the poles is seen, unlike the Stokes V images, and also where the models suggest the strongest magnetic flux should appear in the axisymmetric dipolar field mode. Since our reconstructions require \ispot/\iphot\ $=$\ $0.26$ (see \S \ref{section:results_gj65a}, \ref{section:results_gj65b} \& Fig. \ref{fig:image_gj65}), the contribution towards Stokes V from the spot regions will be small compared with the polar regions. Hence the Stokes V images are mainly sensitive to the less spotted photosphere regions at the spot edges, latitudes above $\theta = 70$\degs, and at the phases where no spots are recovered. \citet{morin08v374peg} similarly found no polar spot, but low-intermediate latitude spots only in brightness images of the M4V star, V347 Peg, though the spots are somewhat weaker than seen here for \gjsfb. The accompanying Stokes V images revealed a strong axisymmetric, large scale poloidal field. The magnetic maps may thus not be offering a true picture of the magnetic field geometry, despite appearing to confirm model predictions. Unbiased observation of the magnetic polarity via Stokes V in both unspotted and spotted regions would be desirable and may soon be realised at near infrared wavelengths with upcoming instrumentation \citep{artigau14spirou,lockhart14crires} where photosphere-spot contrasts are expected to be lower.

The magnetic field phase variability seen in Stokes I \gjsfb by \citet{kochulov17} is likely to arise the phase dependent mean spot filling effects we find in Fig. \ref{fig:image_gj65} (bottom right panel), where $<$$\bar{f}_{\phi}$$> = 0.035 \pm 0.027$ with maximum amplitude variability, $\bar{f}_{\phi,{\rm amp}} =  0.12$. We also note that for \gjsfa, \citet{kochulov17} do not see variability in Stokes I. Similarly, from Fig. \ref{fig:image_gj65} (bottom left panel) we find the phase dependent modulation in \gjsfa is less than seen in \gjsfb where $<$$\bar{f}_{\phi}$$> = 0.013 \pm 0.009$, with maximum amplitude variability, $\bar{f}_{\phi,{\rm amp}} =  0.047$ (i.e. 3 times smaller absolute variability and 2.6 times smaller maximum amplitude spot filling variability). For \gjsn (Fig. \ref{fig:image_gj791}, lower panel), $<$$\bar{f}_{\phi}$$> = 0.018 \pm 0.012$ and $\bar{f}_{\phi,{\rm amp}} =  0.053$ (2014 September 03 and 06) and $<$$\bar{f}_{\phi}$$> = 0.015 \pm 0.010$ and $\bar{f}_{\phi,{\rm amp}} = 0.050$ (2015 September 25 and 28). In this respect, \gjsn more closely resembles \gjsfa, although the modulation is stronger (a factor of 2.3 times smaller compared with \gjsfb). We are investigating red optical lines that are particularly sensitive to Zeeman-broadening (Shulyak et al., In prep), and which offer the best change of measuring magnetic field strength modulation due to stellar rotation. This will enable a higher cadence comparison with our targets than the study by \citet{kochulov17}, with the added benefit of affording a direct, simultaneous comparison with our images.

%simulated magnetic fields in fully convective stars and found that a dipolar field structure can be created, explaining the large field strengths measured in some M stars. It was suggested that convection is then responsible for shredding the field, leading to distributed magnetic field across the surface. It is not clear how the recovered spots and spot stability relate to the simulations of \citet{yadav15}. Inevitably, small-scale structure is not recovered by Doppler imaging with only finite resolution, but the stability over the 3 day gap between observations may be used to constrain estimates of field strength and/or convective efficiency on modifying the morphology (i.e. grow and decay rates) of spots.

%\subsection{Differential rotation trends}
%\protect\label{section:diffrot_trends}

\subsection{Differential rotation}
\protect\label{section:discussion_dr}

\begin{figure}
 \begin{center}
  \begin{tabular}{c}
    \hspace{-4mm} 
    \includegraphics[width=80mm,angle=270]{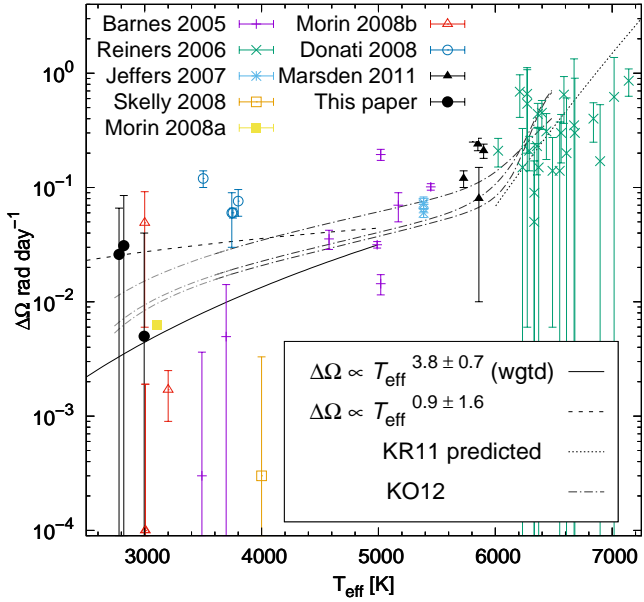}  \\

  \end{tabular}
 \end{center}
 \caption{Differential rotation measured directly using the Doppler imaging sheared image method and Fourier analysis of spectral lines. $\Delta\Omega$ is plotted as a function of effective stellar temperature, $T_{\rm eff}$, for pre-main sequence and main sequence stars. The results from this paper (filled black circles) have been added to measurements made and referenced in \citet{barnes05diffrot} and subsequently in \citet{jeffers07abdor}, \citet{skelly08twa6}, \citet{morin08v374peg}, \citet{donati08mdwarfs}, \citet{morin10mdwarfs}, \citet{marsden11hd141943} and \citet{reiners06diffrot}. For \hbox{$T_{\rm eff} < 5000 $ K}, a power law with \hbox{$\Delta\Omega \propto T_{\rm eff}^{\,3.8 \pm 0.7}$} is found when fitting uncertainties on individual measurements (solid curve). An unweighted fit yields \hbox{$\Delta\Omega \propto T_{\rm eff}^{\,0.9 \pm 1.6}$} (dashed curve). The dotted curve is the predicted relationship for \hbox{$T_{\rm eff} > 6000$} from \citet{kueker11diffrot} of \hbox{$\Delta\Omega \propto T_{\rm eff}^{\,20}$}. The model (dot-dashed) curves from \citet{kitchatinov12} are plotted for stellar rotation periods, $P_{\rm rot}$ = 0.25, 1 and 10 d (bottom to top).}
% \caption{Differential rotation measured directly using the Doppler imaging sheared image method and Fourier analysis of spectral lines. $\Delta\Omega$ is plotted as a function of effective stellar temperature, $T_{\rm eff}$, for pre-main sequence and main sequence stars. The results from this paper (filled black circles) have been added to measurements made and referenced in \citet{barnes05diffrot} and subsequently in \citet{jeffers07abdor}, \citet{skelly08twa6}, \citet{morin08v374peg}, \citet{donati08mdwarfs}, \citet{morin10mdwarfs}, \citet{marsden11hd141943} and \citet{reiners06diffrot}. A power law with \hbox{$\Delta\Omega \propto T_{\rm eff}^{\,4.9 \pm 0.5}$} is found when considering uncertainties on individual measurements (solid curve). An unweighted fit yields \hbox{$\Delta\Omega \propto T_{\rm eff}^{\,5.7 \pm 1.3}$} (dotted curve). The model (dot-dashed) curves from \citet{kitchatinov12} are plotted for stellar rotation periods, $P_{\rm rot}$ = 0.25, 1 and 10 d (bottom to top).}
\protect\label{fig:image_dr_temp}
\end{figure}

In \hbox{Fig. \ref{fig:image_dr_temp}}, we plot $\Delta\Omega$ against $T_{\rm eff}$ for \gjsn, \gjsfa and \gjsfb along with the measurements made spectroscopically for all F, G, K and M dwarfs. Estimates of $\Delta\Omega$ for LQ Lup and R58 that we used in \citet{barnes05diffrot} have been replaced by those from \citet{marsden11hd141943}. Following \citet{cameron07diffrot}, we also use the AB Dor measurements made in \citet{jeffers07abdor}. \citet{barnes05diffrot} found a simple power law relationship, with $\Delta\Omega \propto T_{\rm eff}^{8.9}$. This was subsequently revisited by \citet{reiners06diffrot}, incorporating results for F dwarfs using Fourier analysis. \citet{cameron07diffrot} further revised the power law fit to include subsequent measurements, finding \hbox{$\Delta\Omega \propto T_{\rm eff}^{8.6}$}. \citet{kueker11diffrot} however found it is not appropriate to fit a single power law to the entire temperature range. Their theoretically derived relationship can be split into two regions, with a weaker power law for cooler stars, with $\Delta\Omega \propto T_{\rm eff}^2$ for \hbox{$3800$\ K $< T_{\rm eff} < 5000$\ K}, and a much stronger power law of $\Delta\Omega \propto T_{\rm eff}^{\,20}$ for \hbox{$T_{\rm eff} > 6000$\ K}. Similarly \citet{kitchatinov12}(KO12) derived an analytical relationship for different stellar models. We find that \hbox{$\Delta\Omega = (0.045 \pm 0.013)\,(T_{\rm eff}/5500)^{\,3.8\pm0.7}$} for \hbox{$T_{\rm eff} < 5000$ K} when the uncertainties on each $\Delta\Omega$ value are used. This considerably weaker dependence compared with previous measurements is the result of relative large scatter and exclusion of higher $\Delta\Omega$ values at \hbox{$T_{\rm eff} > 5000$\ K} and is in closer agreement with the $T_{\rm eff}^2$ relationship predicted by \citet{kueker11diffrot}. A fit that excludes error weighting is also shown in Fig. \ref{fig:image_dr_temp}, with \hbox{$\Delta\Omega = (0.045\pm0.027)\,(T_{\rm eff}/5500)^{\,0.9\pm1.6}$}. Since the points with \hbox{$T_{\rm eff} > 6000$\ K} measured by \citet{reiners06diffrot} have large uncertainties and span a relatively large range, we have plotted the prediction of \citet{kueker11diffrot} as a dashed curve (we find \hbox{$\Delta\Omega = (0.085 \pm 0.032)\,(T_{\rm eff}/5500)^{\,7.7\pm2.2}$}). 

%% DIFFERENTIAL ROTATION POWER LAW PLOT

The parameterised relationships found by \citet{kitchatinov12} are also plotted as dash-dot curves in Fig. \ref{fig:image_dr_temp} for three different rotation rates. Curves for \hbox{$P_{\rm rot} =$} \hbox{$0.5, 1$ and $10$ d} are shown. The models are only valid for \hbox{$P_{\rm rot} \gtrsim 0.75$ d}, hence the \hbox{$P_{\rm rot} = 0.5$ d} curve should be treated with caution; however the dependence on rotation { was found to be} relatively weak as also found empirically by \citealt{barnes05diffrot} and more recently by \citet{reinhold15diffrot}\ and\ \citet{balona16a} from Kepler lightcurve studies. In addition, the curves are extrapolated below $M_* = 0.5$ \msun ($T_{\rm eff}$ = 3600 K) where they are plotted in a lighter grey. We have not attempted a numerical comparison of the models since there is considerable scatter. In addition, the stars at the higher temperature end generally have longer rotation periods. The rotation periods of most of the stars in the sample are $< 1$ d, although for HD 141943 \citep{marsden11hd141943}, $P_{\rm rot} = 2.8$ d and the \citet{reiners06diffrot} sample possess minimum periods of \hbox{$P_{\rm rot}/{\rm sin}\,i = $} \hbox{$3.3 \pm 2.0$ d}. In some cases error bars could be underestimated, while it has been shown that the degree of differential rotation for a given object can vary \citep{cameron02twist,jeffers07abdor}. This finding is factored into the multi-epoch measurements for the higher temperature stars reported by \citet{marsden11hd141943}. The dramatic increase noted by \citet{marsden11hd141943} amongst early G stars is evident in the models, though the exact location of this increase may occur at lower temperature than KO12 predict.

%The relatively high $\Delta\Omega$ measurements reported by \citet{donati08mdwarfs} may be significant. The stars in that study possess longer rotation periods of  $P_{\rm rot} = 2.8\,-\,14$ d, so the higher $\Delta\Omega$ may be observational confirmation of the dependence of $\Delta\Omega$ on $\Omega$ predicted by KO12. The measurements for the three stars near the fully convective boundary made by \citet{morin08mdwarfs} would however appear to contradict this argument since the two targets with the smallest $\Delta\Omega$ possess longer periods (\hbox{$P_{\rm rot} = 2.8$ d} and \hbox{$4.4$ d}) compared with the target with the highest $\Delta\Omega$ (\hbox{$P_{\rm rot} = 1.1$ d}). { Suppression of convection by magnetic fields leading to modified convective turnover times, $tau_{\rm c}$, is also likely to lead to modification of differential rotation. Hence the Rosby number, $R_{\rm o} = P_{\rm rot} / tau_{\rm c}$, may provide a tighter correlation with $\Delta\Omega$. The study by \cite{stepian07} demonstrates the correlation for M dwarfs between activity (as measured by the ratio of X-ray to bolometric luminosity, Log(L{\rm x} / L_{\rm bol})) and $R_{\rm o}$ and 

\citet{reinhold13diffrot} and \citet{reinhold15diffrot} took advantage of the large number of stars and extensive Kepler lightcurves to study differential rotation from photometric periodicities. These studies find a large scatter in $\Delta\Omega$ for a given $T_{\rm eff}$ that are likely due to systematic effects of incomplete sampling of lightcurve periodicities. The large number of observations of this type of study enable a statistical comparison with the models and are in good agreement with the predictions of \cite{kueker11diffrot}. Our revised power law is now also in much closer agreement with these studies. Further observational studies by \citet{balona16a} using Kepler lightcurves have enabled empirical parameterisation of $\Delta\Omega$ in terms of both $T_{\rm eff}$ and $\Omega$. Our targets are at the extremes and extend the range most of these studies (i.e. lower $T_{\rm eff}$ and higher $\Omega > 20$ rad day$^{-1}$). Our spectroscopic measurements of $\Delta\Omega$ are nevertheless in good agreement with the extrapolated findings of \citet{balona16a} (Figs. 5 \& 6) and \citet{balona16b} (Figs. 3 \& 4).

\citet{gastine13} and \citet{yadav15} find that although differential rotation is expected to be small in stars showing axisymmetric dipolar fields, significant rotation may be found in those objects that display more multipolar fields. \citet{gastine13} find that $\Delta\Omega / \Omega \sim 5$ per cent for multipolar fields. \citet{yadav15} predict $\Delta\Omega/ \Omega \sim 2$ per cent and note that this is consistent with observations of \citet{morin08mdwarfs}. In fact the four measurements made by \citet{morin08v374peg} and \citet{morin08mdwarfs} yield $\Delta\Omega/ \Omega = 0.02, 0.05, 0.1$\ and\ $0.7$ per cent (i.e. somewhat less than the 2 per cent predicted by \citealt{yadav15}). \citet{davenport15gj1243} finds $\Delta\Omega/ \Omega = 0.1$ per cent from photometric modelling of Kepler data of GJ 1243 (M4V). Similarly, for \gjsn, \gjsfa and \gjsfb, we obtain respective estimates of $\Delta\Omega/ \Omega = 0.02 \pm 0.17$, $0.12 \pm 0.21$ and $0.08 \pm 0.14$ per cent, an order of magnitude lower than predicted by \citealt{yadav15}. Estimates of $\Delta\Omega/ \Omega > 2$ per cent are only found amongst M dwarfs above the fully convective boundary using the sheared image method \citep{donati08mdwarfs} and photometric analysis.

{ Magnetic field strengths and topologies are found to differ in M dwarfs \citep{morin10mdwarfs} as a function of the Rosby number, $R_{\rm o} = P_{\rm rot} / \tau_{\rm c}$. The scatter seen in differential rotation measurements for M dwarfs may be a consequence of the impact of magnetic fields on the convective turnover time, $\tau_{\rm c}$, and may go some way to explaining the spread of $\Delta\Omega$ seen in Fig. \ref{fig:image_dr_temp}. Most of the more slowly rotating M dwarf measurements by \citet{donati08mdwarfs} and \citet{morin08mdwarfs} are also made from Stokes V imaging with spherical harmonic constraints. It has been suggested that the differential rotation measurements recovered using Stokes V profiles are often higher because the magnetic features probe a higher part of the convection zone compared with the more deeply anchored cool spots. It is unclear whether this argument applies to fully convective stars and the Stokes V measurements reported by \citet{donati08mdwarfs} and \citet{morin08mdwarfs}. Unfortunately, those stars rotate too slowly for differential rotation measurements from brightness imaging. Measurement systematics may also be important: the { customary} means of treating errors in Doppler imaging, using \chisq and formal errors, likely underestimates the uncertainties. The Stokes V images of these low-mass M dwarfs relies on fewer profiles and more model assumptions than results from Stokes I. Though computationally expensive, it would be useful to attempt numerical simulations using to Monte Carlo techniques to derive uncertainties for all targets in a consistent manner.}

%The magnetic field strengths and topologies were as a function of the Rosby number, $R_{\rm o} = P_{\rm rot} / tau_{\rm c}$, were investigated by \cite{morin10mdwarfs}. Different saturation levels are seen for high and intermediate mass stars, while the lowest mass stars are seen to 

%The differences in magnetic field strengths measured amongst M dwarfs may impact on the 

%convective turnover time, $tau_{\rm c}$.

%Hence the Rosby number, $R_{\rm o} = P_{\rm rot} / tau_{\rm c}$, is likely to be an important parameter The magnetic field strengths and dynamo bi-stability was investigated as a function of Rosby number

%for fully convective stars with low Rosby numbers was investigated by 

% DeltaOmega/Omega
% Morin 2008 : EQ Peg ( 0.73% ) , YZ Cmi ( 0.017% ) , EV Lac ( 0.12% ) , V374Tau (0.045%)
% This paper :
%gnuplot> print 100*0.005/(2*pi/0.30847),sqrt( (0.035/0.005)**2 + (0.00053/0.30847)**2)*100*0.005/(2*pi/0.30847)
%0.0245472626477769 0.171830843710517
%gnuplot> print 100*0.031/(2*pi/0.24297),sqrt( (0.054/0.031)**2 + (0.00051/0.24297)**2)*100*0.031/(2*pi/0.24297)  
%0.119876617221409 0.208817484827219
%gnuplot> print 100*0.026/(2*pi/0.22679),sqrt( (0.040/0.026)**2 + (0.00033/0.22679)**2)*100*0.026/(2*pi/0.22679)
%0.0938463488139023 0.144379062752659

\section{Concluding remarks}
\protect\label{section:conclusion}

%High resolution spectroscopy reveals that fully convective stars exhibit complex spot patterns. While rapidly rotating G and K stars show similarities and differences in spot patterns to the more sedately rotating Sun, we do not have a fully resolved, fully convective M dwarf with which to compare images. Instead, existing optical instrumentation, such as {\sc uves}, operating at high resolution offers the means to probe more moderately rotating M dwarfs.

With this work and other pre-cursors we have taken the first step towards realising for M dwarfs a fully resolved, fully convective M dwarf with which to compare model images. The high throughput of {\sc uves} operating at {\sc vlt} has enabled high resolution Doppler images of the faint prototype flare star, \gjsfb (UV Cet), to be obtained for the first time. Spot patterns that contrast with its twin, \gjsfa and \gjsn suggest the dynamo mechanism operating in these stars may take different forms, confirming results from magnetic imaging and modelling studies. 
%While rapidly rotating G and K stars show similarities and differences in spot patterns to the more sedately rotating Sun, 

The clarity of signals that we have found suggests that existing optical instrumentation, such as {\sc uves}, operating at high resolution also offers the
means to probe much more moderately rotating M dwarfs. \gjsfa/\gjsfb analogues with more moderate rotation of \vsini \hbox{$\sim 10\,-\,15$ \kms} are expected to rotate with \hbox{$\sim 0.5\,-\,1.0$ d} periods for instance. A narrower 0.3 arcsec slit (yielding \hbox{$R \sim 110,000$} in the red arm of {\sc uves}) and the reduced need for very short exposures mean that stars with \hbox{I $\sim 12$} (i.e. $\sim 3$ magnitudes fainter than the targets studied in this paper) can be imaged with spectra of comparable quality. Deeper lines relative to the normalised continuum would offset the reduced S/N $\sim 50\,-\,70$ (assuming $\sim$$900$ sec exposures) we would expect in the extracted spectra, while at \hbox{\vsini\ $\sim 15$ \kms} and \hbox{$R \sim 110,000$}, the effective spot resolution would be $\sim 0.6$ of what is achieved here.

Near infrared technology covering multiple orders at high resolution \citep{lizon14crires,artigau14spirou} will also enable us to investigate wavelength dependent starspot contrast effects on a fainter sample of stars. With improvements in theoretical line lists, it may also be possible to investigate individual molecular species or transition energy ranges. Infrared polarimetry will be particularly important for assessing the optical Stokes V observations, which do not enable reliable estimation of the magnetic field inside the spots we recover with brightness imaging. In addition to providing information about the underlying dynamo mechanisms, an understanding of starspot distributions on fully convective stars, which may possess significant rotation on average (with \vsini\ $\sim 10$\ \kms, \citealt{jenkins09mdwarfs,reiners12rotation}), is needed if we are to effectively deal with radial velocity jitter in radial velocity searches for planets \citep{barnes17jitter}. Both CRIRES+ and and {\sc uves} will be important instruments in this respect, offering the opportunity for high resolution observations.

%.. speculate on what range of late type stars could be looked at before signal strength /phase coverage becomes an issue, perhaps you could look at the line list, maybe it is possible with more data to produce spot patterns for particular species .. sensitivity to different contrast spots at different latitudes, don’t think I am aware of any models for this to make interesting predictions, simply, forthcoming telescope proposal! More importantly perhaps do you have an hypothesis to to explain what is going on?

\section*{Acknowledgments}
%{We thank the anonymous referee for taking the time to review this manuscript.} 
{ We would like to thank the referee, Pascal Petit, for taking the time to read the manuscript and for constructive comments that helped improve the paper.}
This work is based on observations collected at the European Organisation for Astronomical Research in the Southern Hemisphere under ESO programmes 093.D-0165(A) and 095.D-0291(A). J.R.B. and C.A.H. were supported by the STFC under the grant ST/L000776/1 { and ST/P000584/1}. S.V.J. acknowledges research funding by the Deutsche Forschungsgemeinschaft (DFG) under grant SFB 963/1, project A16. HJ was supported by grants from the Leverhulme Trust (RPG-2014-281) and the Science and Technology Facilities Council (ST/M001008/1). JSJ acknowledges funding by Fondecyt through grants 1161218 and 3110004, and partial support from CATA-Basal (PB06, Conicyt), the GEMINI-CONICYT FUND and from the Comit\'{e} Mixto ESO-GOBIERNO DE CHILE. 

%\bibliographystyle{mnras}
%\bibliography{iau_journals,master,ownrefs,planets}

\begin{thebibliography}{}
\makeatletter
\relax
\def\mn@urlcharsother{\let\do\@makeother \do\$\do\&\do\#\do\^\do\_\do\%\do\~}
\def\mn@doi{\begingroup\mn@urlcharsother \@ifnextchar [ {\mn@doi@}
  {\mn@doi@[]}}
\def\mn@doi@[#1]#2{\def\@tempa{#1}\ifx\@tempa\@empty \href
  {http://dx.doi.org/#2} {doi:#2}\else \href {http://dx.doi.org/#2} {#1}\fi
  \endgroup}
\def\mn@eprint#1#2{\mn@eprint@#1:#2::\@nil}
\def\mn@eprint@arXiv#1{\href {http://arxiv.org/abs/#1} {{\tt arXiv:#1}}}
\def\mn@eprint@dblp#1{\href {http://dblp.uni-trier.de/rec/bibtex/#1.xml}
  {dblp:#1}}
\def\mn@eprint@#1:#2:#3:#4\@nil{\def\@tempa {#1}\def\@tempb {#2}\def\@tempc
  {#3}\ifx \@tempc \@empty \let \@tempc \@tempb \let \@tempb \@tempa \fi \ifx
  \@tempb \@empty \def\@tempb {arXiv}\fi \@ifundefined
  {mn@eprint@\@tempb}{\@tempb:\@tempc}{\expandafter \expandafter \csname
  mn@eprint@\@tempb\endcsname \expandafter{\@tempc}}}

\bibitem[\protect\citeauthoryear{{Allard}, {Homeier}, {Freytag}  \&
  {Sharp}}{{Allard} et~al.}{2012}]{allard12}
{Allard} F.,  {Homeier} D.,  {Freytag} B.,   {Sharp} C.~M.,  2012, in
  {Reyl{\'e}} C.,  {Charbonnel} C.,   {Schultheis} M.,  eds,  EAS Publications
  Series Vol. 57, EAS Publications Series. pp 3--43 (\mn@eprint {arXiv}
  {1206.1021}), \mn@doi{10.1051/eas/1257001}

\bibitem[\protect\citeauthoryear{{Artigau} et~al.,}{{Artigau}
  et~al.}{2014}]{artigau14spirou}
{Artigau} {\'E}.,  et~al., 2014, in Society of Photo-Optical Instrumentation
  Engineers (SPIE) Conference Series. p.~15 (\mn@eprint {arXiv} {1406.6992}),
  \mn@doi{10.1117/12.2055663}

\bibitem[\protect\citeauthoryear{{Balona} \& {Abedigamba}}{{Balona} \&
  {Abedigamba}}{2016}]{balona16a}
{Balona} L.~A.,  {Abedigamba} O.~P.,  2016, \mn@doi [MNRAS]
  {10.1093/mnras/stw1443}, \href
  {http://cdsads.u-strasbg.fr/abs/2016MNRAS.461..497B} {461, 497}

\bibitem[\protect\citeauthoryear{{Balona}, {{\v S}vanda}  \&
  {Karlick{\'y}}}{{Balona} et~al.}{2016}]{balona16b}
{Balona} L.~A.,  {{\v S}vanda} M.,   {Karlick{\'y}} M.,  2016, \mn@doi [MNRAS]
  {10.1093/mnras/stw2109}, \href
  {http://cdsads.u-strasbg.fr/abs/2016MNRAS.463.1740B} {463, 1740}

\bibitem[\protect\citeauthoryear{{Baraffe}, {Homeier}, {Allard}  \&
  {Chabrier}}{{Baraffe} et~al.}{2015}]{baraffe15}
{Baraffe} I.,  {Homeier} D.,  {Allard} F.,   {Chabrier} G.,  2015, \mn@doi
  [A\&A] {10.1051/0004-6361/201425481}, \href
  {http://cdsads.u-strasbg.fr/abs/2015A%26A...577A..42B} {577, A42}

\bibitem[\protect\citeauthoryear{{Barnes} \& {Collier Cameron}}{{Barnes} \&
  {Collier Cameron}}{2001}]{barnes01mdwarfs}
{Barnes} J.~R.,  {Collier Cameron} A.,  2001, MNRAS, 326, 950

\bibitem[\protect\citeauthoryear{{Barnes}, {Collier Cameron}, {Unruh}, {Donati}
   \& {Hussain}}{{Barnes} et~al.}{1998}]{barnes98aper}
{Barnes} J.~R.,  {Collier Cameron} A.,  {Unruh} Y.~C.,  {Donati} J.~F.,
  {Hussain} G.~A.~J.,  1998, MNRAS, 299, 904

\bibitem[\protect\citeauthoryear{{Barnes}, {Collier Cameron}, {James}  \&
  {Steeghs}}{{Barnes} et~al.}{2001}]{barnes01aper}
{Barnes} J.~R.,  {Collier Cameron} A.,  {James} D.~J.,   {Steeghs} D.,  2001,
  MNRAS, 326, 1057

\bibitem[\protect\citeauthoryear{{Barnes}, {James}  \& {Cameron}}{{Barnes}
  et~al.}{2004}]{barnes04hkaqr}
{Barnes} J.~R.,  {James} D.~J.,   {Cameron} A.~C.,  2004, MNRAS, 352, 589

\bibitem[\protect\citeauthoryear{{Barnes}, {Cameron}, {Donati}, {James},
  {Marsden}  \& {Petit}}{{Barnes} et~al.}{2005}]{barnes05diffrot}
{Barnes} J.~R.,  {Cameron} A.~C.,  {Donati} J.-F.,  {James} D.~J.,  {Marsden}
  S.~C.,   {Petit} P.,  2005, MNRAS, 357, L1

\bibitem[\protect\citeauthoryear{{Barnes} et~al.,}{{Barnes}
  et~al.}{2012}]{barnes12rops}
{Barnes} J.~R.,  et~al., 2012, \mn@doi [MNRAS]
  {10.1111/j.1365-2966.2012.21236.x}, \href
  {http://cdsads.u-strasbg.fr/abs/2012MNRAS.424..591B} {424, 591}

\bibitem[\protect\citeauthoryear{{Barnes}, {Jeffers}, {Jones}, {Pavlenko},
  {Jenkins}, {Haswell}  \& {Lohr}}{{Barnes} et~al.}{2015}]{barnes15mdwarfsB15}
{Barnes} J.~R.,  {Jeffers} S.~V.,  {Jones} H.~R.~A.,  {Pavlenko} Y.~V.,
  {Jenkins} J.~S.,  {Haswell} C.~A.,   {Lohr} M.~E.,  2015, \mn@doi [ApJ]
  {10.1088/0004-637X/812/1/42}, \href
  {http://ukads.nottingham.ac.uk/abs/2015ApJ...812...42B} {812, 42 (B15)}

\bibitem[\protect\citeauthoryear{{Barnes} et~al.,}{{Barnes}
  et~al.}{2017}]{barnes17jitter}
{Barnes} J.~R.,  et~al., 2017, \mn@doi [MNRAS] {10.1093/mnras/stw3170}, \href
  {http://cdsads.u-strasbg.fr/abs/2017MNRAS.466.1733B} {466, 1733}

\bibitem[\protect\citeauthoryear{{Benedict}, {McArthur}, {Franz}, {Wasserman}
  \& {Henry}}{{Benedict} et~al.}{2000}]{benedict00gj791}
{Benedict} G.~F.,  {McArthur} B.~E.,  {Franz} O.~G.,  {Wasserman} L.~H.,
  {Henry} T.~J.,  2000, \mn@doi [AJ] {10.1086/301495}, \href
  {http://cdsads.u-strasbg.fr/abs/2000AJ....120.1106B} {120, 1106}

\bibitem[\protect\citeauthoryear{{Benedict} et~al.,}{{Benedict}
  et~al.}{2016}]{benedict16}
{Benedict} G.~F.,  et~al., 2016, \mn@doi [AJ] {10.3847/0004-6256/152/5/141},
  \href {http://cdsads.u-strasbg.fr/abs/2016AJ....152..141B} {152, 141}

\bibitem[\protect\citeauthoryear{{Berdyugina}}{{Berdyugina}}{2005}]{berdyugina05starspots}
{Berdyugina} S.~V.,  2005, \mn@doi [Living Reviews in Solar Physics]
  {10.12942/lrsp-2005-8}, \href
  {http://cdsads.u-strasbg.fr/abs/2005LRSP....2....8B} {2, 8}

\bibitem[\protect\citeauthoryear{{Bopp} \& {Moffett}}{{Bopp} \&
  {Moffett}}{1973}]{bopp73gj65}
{Bopp} B.~W.,  {Moffett} T.~J.,  1973, \mn@doi [ApJ] {10.1086/152412}, \href
  {http://cdsads.u-strasbg.fr/abs/1973ApJ...185..239B} {185, 239}

\bibitem[\protect\citeauthoryear{{Brandenburg}}{{Brandenburg}}{2005}]{brandenburg05dynamo}
{Brandenburg} A.,  2005, \mn@doi [ApJ] {10.1086/429584}, \href
  {http://cdsads.u-strasbg.fr/abs/2005ApJ...625..539B} {625, 539}

\bibitem[\protect\citeauthoryear{{Browning}}{{Browning}}{2008}]{browning08}
{Browning} M.~K.,  2008, \mn@doi [ApJ] {10.1086/527432}, \href
  {http://cdsads.u-strasbg.fr/abs/2008ApJ...676.1262B} {676, 1262}

\bibitem[\protect\citeauthoryear{{Brun} et~al.,}{{Brun}
  et~al.}{2017}]{brun17diffrot}
{Brun} A.~S.,  et~al., 2017, \mn@doi [ApJ] {10.3847/1538-4357/aa5c40}, \href
  {http://cdsads.u-strasbg.fr/abs/2017ApJ...836..192B} {836, 192}

\bibitem[\protect\citeauthoryear{{Chabrier}, {Gallardo}  \&
  {Baraffe}}{{Chabrier} et~al.}{2007}]{chabrier07}
{Chabrier} G.,  {Gallardo} J.,   {Baraffe} I.,  2007, \mn@doi [A\&A]
  {10.1051/0004-6361:20077702}, \href
  {http://cdsads.u-strasbg.fr/abs/2007A%26A...472L..17C} {472, L17}

\bibitem[\protect\citeauthoryear{Collier~Cameron}{Collier~Cameron}{2001}]{cameron01mapping}
Collier~Cameron A.,  2001, in {{Boffin}, H.~M.~J. and {Steeghs}, D. and
  {Cuypers}, J.} ed., Astrotomography - Indirect Imaging Methods in
  Observational Astronomy. Springer (Lecture Notes in Physics), pp 183--206

\bibitem[\protect\citeauthoryear{{Collier Cameron}}{{Collier
  Cameron}}{2007}]{cameron07diffrot}
{Collier Cameron} A.,  2007, \mn@doi [Astronomische Nachrichten]
  {10.1002/asna.200710880}, \href
  {http://cdsads.u-strasbg.fr/abs/2007AN....328.1030C} {328, 1030}

\bibitem[\protect\citeauthoryear{{Collier Cameron} \& {Donati}}{{Collier
  Cameron} \& {Donati}}{2002}]{cameron02twist}
{Collier Cameron} A.,  {Donati} J.-F.,  2002, MNRAS, 329, L23

\bibitem[\protect\citeauthoryear{{DENIS Consortium}}{{DENIS
  Consortium}}{2003}]{denis03}
{DENIS Consortium} 2003, VizieR Online Data Catalog, \href
  {http://cdsads.u-strasbg.fr/abs/2003yCat.2252....0D} {2252}

\bibitem[\protect\citeauthoryear{{Davenport}, {Hebb}  \& {Hawley}}{{Davenport}
  et~al.}{2015}]{davenport15gj1243}
{Davenport} J.~R.~A.,  {Hebb} L.,   {Hawley} S.~L.,  2015, \mn@doi [ApJ]
  {10.1088/0004-637X/806/2/212}, \href
  {http://cdsads.u-strasbg.fr/abs/2015ApJ...806..212D} {806, 212}

\bibitem[\protect\citeauthoryear{{De Pontieu}, {Carlsson}, {Stein}, {Rouppe van
  der Voort}, {L{\"o}fdahl}, {van Noort}, {Nordlund}  \& {Scharmer}}{{De
  Pontieu} et~al.}{2006}]{depontieu06}
{De Pontieu} B.,  {Carlsson} M.,  {Stein} R.,  {Rouppe van der Voort} L.,
  {L{\"o}fdahl} M.,  {van Noort} M.,  {Nordlund} {\AA}.,   {Scharmer} G.,
  2006, \mn@doi [ApJ] {10.1086/505074}, \href
  {http://cdsads.u-strasbg.fr/abs/2006ApJ...646.1405D} {646, 1405}

\bibitem[\protect\citeauthoryear{Donati, Semel, Carter, Rees  \&
  Collier~Cameron}{Donati et~al.}{1997}]{donati97zdi}
Donati J.-F.,  Semel M.,  Carter B.,  Rees D.~E.,   Collier~Cameron A.,  1997,
  MNRAS, 291, 658

\bibitem[\protect\citeauthoryear{{Donati} et~al.,}{{Donati}
  et~al.}{2008}]{donati08mdwarfs}
{Donati} J.,  et~al., 2008, \mn@doi [MNRAS] {10.1111/j.1365-2966.2008.13799.x},
  \href {http://cdsads.u-strasbg.fr/abs/2008MNRAS.390..545D} {390, 545}

\bibitem[\protect\citeauthoryear{{Feiden} \& {Chaboyer}}{{Feiden} \&
  {Chaboyer}}{2012}]{feiden12}
{Feiden} G.~A.,  {Chaboyer} B.,  2012, \mn@doi [ApJ]
  {10.1088/0004-637X/757/1/42}, \href
  {http://cdsads.u-strasbg.fr/abs/2012ApJ...757...42F} {757, 42}

\bibitem[\protect\citeauthoryear{{Gadun} \& {Pavlenko}}{{Gadun} \&
  {Pavlenko}}{1997}]{gadun97}
{Gadun} A.~S.,  {Pavlenko} Y.~V.,  1997, \aap, \href
  {http://cdsads.u-strasbg.fr/abs/1997A%26A...324..281G} {324, 281}

\bibitem[\protect\citeauthoryear{{Gastine}, {Duarte}  \& {Wicht}}{{Gastine}
  et~al.}{2012}]{gastine12}
{Gastine} T.,  {Duarte} L.,   {Wicht} J.,  2012, \mn@doi [A\&A]
  {10.1051/0004-6361/201219799}, \href
  {http://cdsads.u-strasbg.fr/abs/2012A%26A...546A..19G} {546, A19}

\bibitem[\protect\citeauthoryear{{Gastine}, {Morin}, {Duarte}, {Reiners},
  {Christensen}  \& {Wicht}}{{Gastine} et~al.}{2013}]{gastine13}
{Gastine} T.,  {Morin} J.,  {Duarte} L.,  {Reiners} A.,  {Christensen} U.~R.,
  {Wicht} J.,  2013, \mn@doi [A\&A] {10.1051/0004-6361/201220317}, \href
  {http://cdsads.u-strasbg.fr/abs/2013A%26A...549L...5G} {549, L5}

\bibitem[\protect\citeauthoryear{{Heavens}}{{Heavens}}{1993}]{heavens93redshifts}
{Heavens} A.~F.,  1993, MNRAS, \href
  {http://cdsads.u-strasbg.fr/abs/1993MNRAS.263..735H} {263, 735}

\bibitem[\protect\citeauthoryear{{Henry}, {Ianna}, {Kirkpatrick}  \&
  {Jahreiss}}{{Henry} et~al.}{1997}]{henry97recons}
{Henry} T.~J.,  {Ianna} P.~A.,  {Kirkpatrick} J.~D.,   {Jahreiss} H.,  1997,
  \mn@doi [\aj] {10.1086/118482}, \href
  {http://cdsads.u-strasbg.fr/abs/1997AJ....114..388H} {114}

\bibitem[\protect\citeauthoryear{Horne}{Horne}{1986}]{horne86extopt}
Horne K.~D.,  1986, PASP, 98, 609

\bibitem[\protect\citeauthoryear{{Hosey}, {Henry}, {Jao}, {Dieterich},
  {Winters}, {Lurie}, {Riedel}  \& {Subasavage}}{{Hosey}
  et~al.}{2015}]{hosey15}
{Hosey} A.~D.,  {Henry} T.~J.,  {Jao} W.-C.,  {Dieterich} S.~B.,  {Winters}
  J.~G.,  {Lurie} J.~C.,  {Riedel} A.~R.,   {Subasavage} J.~P.,  2015, \mn@doi
  [\aj] {10.1088/0004-6256/150/1/6}, \href
  {http://cdsads.u-strasbg.fr/abs/2015AJ....150....6H} {150, 6}

\bibitem[\protect\citeauthoryear{{Jeffers}, {Donati}  \& {Collier
  Cameron}}{{Jeffers} et~al.}{2007}]{jeffers07abdor}
{Jeffers} S.~V.,  {Donati} J.-F.,   {Collier Cameron} A.,  2007, \mn@doi
  [MNRAS] {10.1111/j.1365-2966.2006.11154.x}, \href
  {http://ukads.nottingham.ac.uk/abs/2007MNRAS.375..567J} {375, 567}

\bibitem[\protect\citeauthoryear{{Jenkins}, {Ramsey}, {Jones}, {Pavlenko},
  {Gallardo}, {Barnes}  \& {Pinfield}}{{Jenkins}
  et~al.}{2009}]{jenkins09mdwarfs}
{Jenkins} J.~S.,  {Ramsey} L.~W.,  {Jones} H.~R.~A.,  {Pavlenko} Y.,
  {Gallardo} J.,  {Barnes} J.~R.,   {Pinfield} D.~J.,  2009, \mn@doi [ApJ]
  {10.1088/0004-637X/704/2/975}, \href
  {http://cdsads.u-strasbg.fr/abs/2009ApJ...704..975J} {704, 975}

\bibitem[\protect\citeauthoryear{{Johns-Krull} \& {Valenti}}{{Johns-Krull} \&
  {Valenti}}{1996}]{johnskrull96mdwarfs}
{Johns-Krull} C.~M.,  {Valenti} J.~A.,  1996, \mn@doi [ApJ] {10.1086/309954},
  \href {http://cdsads.u-strasbg.fr/abs/1996ApJ...459L..95J} {459, L95+}

\bibitem[\protect\citeauthoryear{{Kervella}, {M{\'e}rand}, {Ledoux}, {Demory}
  \& {Le Bouquin}}{{Kervella} et~al.}{2016}]{kervella16gj65}
{Kervella} P.,  {M{\'e}rand} A.,  {Ledoux} C.,  {Demory} B.-O.,   {Le Bouquin}
  J.-B.,  2016, \mn@doi [\aap] {10.1051/0004-6361/201628631}, \href
  {http://cdsads.u-strasbg.fr/abs/2016A%26A...593A.127K} {593, A127}

\bibitem[\protect\citeauthoryear{{Kitchatinov} \& {Olemskoy}}{{Kitchatinov} \&
  {Olemskoy}}{2012}]{kitchatinov12}
{Kitchatinov} L.~L.,  {Olemskoy} S.~V.,  2012, \mn@doi [MNRAS]
  {10.1111/j.1365-2966.2012.21126.x}, \href
  {http://cdsads.u-strasbg.fr/abs/2012MNRAS.423.3344K} {423, 3344 (KO12)}

\bibitem[\protect\citeauthoryear{{Kitchatinov}, {Moss}  \&
  {Sokoloff}}{{Kitchatinov} et~al.}{2014}]{kitchatinov14}
{Kitchatinov} L.~L.,  {Moss} D.,   {Sokoloff} D.,  2014, \mn@doi [MNRAS]
  {10.1093/mnrasl/slu041}, \href
  {http://cdsads.u-strasbg.fr/abs/2014MNRAS.442L...1K} {442, L1}

\bibitem[\protect\citeauthoryear{{Kochukhov} \& {Lavail}}{{Kochukhov} \&
  {Lavail}}{2017}]{kochulov17}
{Kochukhov} O.,  {Lavail} A.,  2017, \mn@doi [ApJ]
  {10.3847/2041-8213/835/1/L4}, \href
  {http://cdsads.u-strasbg.fr/abs/2017ApJ...835L...4K} {835, L4}

\bibitem[\protect\citeauthoryear{{K{\"u}ker} \& {R{\"u}diger}}{{K{\"u}ker} \&
  {R{\"u}diger}}{2011}]{kueker11diffrot}
{K{\"u}ker} M.,  {R{\"u}diger} G.,  2011, \mn@doi [Astronomische Nachrichten]
  {10.1002/asna.201111628}, \href
  {http://cdsads.u-strasbg.fr/abs/2011AN....332..933K} {332, 933}

\bibitem[\protect\citeauthoryear{{Lizon} et~al.,}{{Lizon}
  et~al.}{2014}]{lizon14crires}
{Lizon} J.~L.,  et~al., 2014, in Ground-based and Airborne Instrumentation for
  Astronomy V. p. 91477S, \mn@doi{10.1117/12.2054800}

\bibitem[\protect\citeauthoryear{{Lockhart} et~al.,}{{Lockhart}
  et~al.}{2014}]{lockhart14crires}
{Lockhart} M.,  et~al., 2014, in Ground-based and Airborne Instrumentation for
  Astronomy V. p. 91478P, \mn@doi{10.1117/12.2056367}

\bibitem[\protect\citeauthoryear{{Luyten}}{{Luyten}}{1949}]{luyten49gj65}
{Luyten} W.~J.,  1949, \mn@doi [ApJ] {10.1086/145158}, \href
  {http://cdsads.u-strasbg.fr/abs/1949ApJ...109..532L} {109, 532}

\bibitem[\protect\citeauthoryear{{Marsden}, {Mengel}, {Donati}, {Carter},
  {Semel}  \& {Petit}}{{Marsden} et~al.}{2006a}]{marsden06}
{Marsden} S.~C.,  {Mengel} M.~W.,  {Donati} F.,  {Carter} B.~D.,  {Semel} M.,
  {Petit} P.,  2006a, in {Casini} R.,  {Lites} B.~W.,  eds,  Astronomical
  Society of the Pacific Conference Series Vol. 358, Astronomical Society of
  the Pacific Conference Series. p.~401

\bibitem[\protect\citeauthoryear{{Marsden}, {Donati}, {Semel}, {Petit}  \&
  {Carter}}{{Marsden} et~al.}{2006b}]{marsden06hd171488}
{Marsden} S.~C.,  {Donati} J.-F.,  {Semel} M.,  {Petit} P.,   {Carter} B.~D.,
  2006b, \mn@doi [MNRAS] {10.1111/j.1365-2966.2006.10503.x}, \href
  {http://cdsads.u-strasbg.fr/abs/2006MNRAS.370..468M} {370, 468}

\bibitem[\protect\citeauthoryear{{Marsden} et~al.,}{{Marsden}
  et~al.}{2011}]{marsden11hd141943}
{Marsden} S.~C.,  et~al., 2011, \mn@doi [MNRAS]
  {10.1111/j.1365-2966.2011.18367.x}, \href
  {http://ukads.nottingham.ac.uk/abs/2011MNRAS.413.1922M} {413, 1922}

\bibitem[\protect\citeauthoryear{{Mason}, {Wycoff}, {Hartkopf}, {Douglass}  \&
  {Worley}}{{Mason} et~al.}{2001}]{mason01catalogue}
{Mason} B.~D.,  {Wycoff} G.~L.,  {Hartkopf} W.~I.,  {Douglass} G.~G.,
  {Worley} C.~E.,  2001, \mn@doi [AJ] {10.1086/323920}, \href
  {http://adsabs.harvard.edu/abs/2001AJ....122.3466M} {122, 3466}

\bibitem[\protect\citeauthoryear{{Mills}, {Webb}, {Clayton}  \& {Gray}}{{Mills}
  et~al.}{2014}]{mills14echomop}
{Mills} D.,  {Webb} J.,  {Clayton} M.,   {Gray} N.,  2014, {ECHOMOP: Echelle
  data reduction package}, Astrophysics Source Code Library (\mn@eprint {ascl}
  {1405.018})

\bibitem[\protect\citeauthoryear{{Montes}, {L{\'o}pez-Santiago}, {G{\'a}lvez},
  {Fern{\'a}ndez-Figueroa}, {De Castro}  \& {Cornide}}{{Montes}
  et~al.}{2001}]{montes01members}
{Montes} D.,  {L{\'o}pez-Santiago} J.,  {G{\'a}lvez} M.~C.,
  {Fern{\'a}ndez-Figueroa} M.~J.,  {De Castro} E.,   {Cornide} M.,  2001,
  \mn@doi [MNRAS] {10.1046/j.1365-8711.2001.04781.x}, \href
  {http://cdsads.u-strasbg.fr/abs/2001MNRAS.328...45M} {328, 45}

\bibitem[\protect\citeauthoryear{{Moreno-Insertis}, {Sch\"{u}ssler}  \&
  {Ferriz-Mas}}{{Moreno-Insertis} et~al.}{1992}]{moreno92fluxtubes}
{Moreno-Insertis} F.,  {Sch\"{u}ssler} M.,   {Ferriz-Mas} A.,  1992, A\&A, 264,
  686

\bibitem[\protect\citeauthoryear{{Morin} et~al.,}{{Morin}
  et~al.}{2008a}]{morin08v374peg}
{Morin} J.,  et~al., 2008a, \mn@doi [MNRAS] {10.1111/j.1365-2966.2007.12709.x},
  \href {http://cdsads.u-strasbg.fr/abs/2008MNRAS.384...77M} {384, 77}

\bibitem[\protect\citeauthoryear{{Morin} et~al.,}{{Morin}
  et~al.}{2008b}]{morin08mdwarfs}
{Morin} J.,  et~al., 2008b, \mn@doi [MNRAS] {10.1111/j.1365-2966.2008.13809.x},
  \href {http://cdsads.u-strasbg.fr/abs/2008MNRAS.390..567M} {390, 567}

\bibitem[\protect\citeauthoryear{{Morin}, {Donati}, {Petit}, {Delfosse},
  {Forveille}  \& {Jardine}}{{Morin} et~al.}{2010}]{morin10mdwarfs}
{Morin} J.,  {Donati} J.-F.,  {Petit} P.,  {Delfosse} X.,  {Forveille} T.,
  {Jardine} M.~M.,  2010, \mn@doi [MNRAS] {10.1111/j.1365-2966.2010.17101.x},
  \href {http://cdsads.u-strasbg.fr/abs/2010MNRAS.407.2269M} {407, 2269}

\bibitem[\protect\citeauthoryear{{Pavlenko}}{{Pavlenko}}{2014}]{pavlenko14}
{Pavlenko} Y.~V.,  2014, \mn@doi [Astronomy Reports]
  {10.1134/S1063772914110043}, \href
  {http://cdsads.u-strasbg.fr/abs/2014ARep...58..825P} {58, 825}

\bibitem[\protect\citeauthoryear{{Pavlenko} \& {Schmidt}}{{Pavlenko} \&
  {Schmidt}}{2015}]{pavlenko15}
{Pavlenko} Y.~V.,  {Schmidt} M.,  2015, \mn@doi [Kinematics and Physics of
  Celestial Bodies] {10.3103/S0884591315020051}, \href
  {http://cdsads.u-strasbg.fr/abs/2015KPCB...31...90P} {31, 90}

\bibitem[\protect\citeauthoryear{{Pavlenko}, {Jones}, {Mart{\'{\i}}n},
  {Guenther}, {Kenworthy}  \& {Zapatero Osorio}}{{Pavlenko}
  et~al.}{2007}]{pavlenko07lp944}
{Pavlenko} Y.~V.,  {Jones} H.~R.~A.,  {Mart{\'{\i}}n} E.~L.,  {Guenther} E.,
  {Kenworthy} M.~A.,   {Zapatero Osorio} M.~R.,  2007, \mn@doi [MNRAS]
  {10.1111/j.1365-2966.2007.12182.x}, \href
  {http://cdsads.u-strasbg.fr/abs/2007MNRAS.380.1285P} {380, 1285}

\bibitem[\protect\citeauthoryear{{Petit}, {Donati}  \& {Collier
  Cameron}}{{Petit} et~al.}{2002}]{petit02}
{Petit} P.,  {Donati} J.-F.,   {Collier Cameron} A.,  2002, MNRAS, 334, 374

\bibitem[\protect\citeauthoryear{{Phan-Bao}, {Lim}, {Donati}, {Johns-Krull}  \&
  {Mart{\'{\i}}n}}{{Phan-Bao} et~al.}{2009}]{phanbao09}
{Phan-Bao} N.,  {Lim} J.,  {Donati} J.-F.,  {Johns-Krull} C.~M.,
  {Mart{\'{\i}}n} E.~L.,  2009, \mn@doi [ApJ] {10.1088/0004-637X/704/2/1721},
  \href {http://cdsads.u-strasbg.fr/abs/2009ApJ...704.1721P} {704, 1721}

\bibitem[\protect\citeauthoryear{{Reiners}}{{Reiners}}{2006}]{reiners06diffrot}
{Reiners} A.,  2006, \mn@doi [A\&A] {10.1051/0004-6361:20053911}, \href
  {http://ukads.nottingham.ac.uk/abs/2006A%26A...446..267R} {446, 267}

\bibitem[\protect\citeauthoryear{{Reiners} \& {Schmitt}}{{Reiners} \&
  {Schmitt}}{2003}]{reiners03diffrot}
{Reiners} A.,  {Schmitt} J.~H.~M.~M.,  2003, A\&A, 398, 647

\bibitem[\protect\citeauthoryear{{Reiners}, {Joshi}  \& {Goldman}}{{Reiners}
  et~al.}{2012}]{reiners12rotation}
{Reiners} A.,  {Joshi} N.,   {Goldman} B.,  2012, \mn@doi [AJ]
  {10.1088/0004-6256/143/4/93}, \href
  {http://cdsads.u-strasbg.fr/abs/2012AJ....143...93R} {143, 93}

\bibitem[\protect\citeauthoryear{{Reinhold} \& {Gizon}}{{Reinhold} \&
  {Gizon}}{2015}]{reinhold15diffrot}
{Reinhold} T.,  {Gizon} L.,  2015, \mn@doi [A\&A]
  {10.1051/0004-6361/201526216}, \href
  {http://cdsads.u-strasbg.fr/abs/2015A%26A...583A..65R} {583, A65}

\bibitem[\protect\citeauthoryear{{Reinhold}, {Reiners}  \& {Basri}}{{Reinhold}
  et~al.}{2013}]{reinhold13diffrot}
{Reinhold} T.,  {Reiners} A.,   {Basri} G.,  2013, \mn@doi [A\&A]
  {10.1051/0004-6361/201321970}, \href
  {http://cdsads.u-strasbg.fr/abs/2013A%26A...560A...4R} {560, A4}

\bibitem[\protect\citeauthoryear{{Saar} \& {Linsky}}{{Saar} \&
  {Linsky}}{1985}]{saar85adleo}
{Saar} S.~H.,  {Linsky} J.~L.,  1985, \mn@doi [ApJ] {10.1086/184578}, \href
  {http://ukads.nottingham.ac.uk/abs/1985ApJ...299L..47S} {299, L47}

\bibitem[\protect\citeauthoryear{{Sch\"{u}ssler}, {Caligari}, {Ferriz-Mas},
  {Solanki}  \& {Stix}}{{Sch\"{u}ssler} et~al.}{1996}]{schussler96buoy}
{Sch\"{u}ssler} M.,  {Caligari} P.,  {Ferriz-Mas} A.,  {Solanki} S.~K.,
  {Stix} M.,  1996, A\&A, 314, 503

\bibitem[\protect\citeauthoryear{{Shortridge}}{{Shortridge}}{1993}]{shortridge93figaro}
{Shortridge} K.,  1993, in ASP Conf. Ser. 52: Astronomical Data Analysis
  Software and Systems II. pp 219--+

\bibitem[\protect\citeauthoryear{{Shulyak}, {Sokoloff}, {Kitchatinov}  \&
  {Moss}}{{Shulyak} et~al.}{2015}]{shulyak15dynamo}
{Shulyak} D.,  {Sokoloff} D.,  {Kitchatinov} L.,   {Moss} D.,  2015, \mn@doi
  [MNRAS] {10.1093/mnras/stv585}, \href
  {http://cdsads.u-strasbg.fr/abs/2015MNRAS.449.3471S} {449, 3471}

\bibitem[\protect\citeauthoryear{{Skelly}, {Unruh}, {Collier Cameron},
  {Barnes}, {Donati}, {Lawson}  \& {Carter}}{{Skelly}
  et~al.}{2008}]{skelly08twa6}
{Skelly} M.~B.,  {Unruh} Y.~C.,  {Collier Cameron} A.,  {Barnes} J.~R.,
  {Donati} J.-F.,  {Lawson} W.~A.,   {Carter} B.~D.,  2008, \mn@doi [MNRAS]
  {10.1111/j.1365-2966.2008.12917.x}, \href
  {http://cdsads.u-strasbg.fr/abs/2008MNRAS.385..708S} {385, 708}

\bibitem[\protect\citeauthoryear{{Strassmeier}}{{Strassmeier}}{2009}]{strassmeier09starspots}
{Strassmeier} K.~G.,  2009, \mn@doi [The Astronomy and Astrophysics Review]
  {10.1007/s00159-009-0020-6}, \href
  {http://ukads.nottingham.ac.uk/abs/2009A%26ARv..17..251S} {17, 251}

\bibitem[\protect\citeauthoryear{{Yadav}, {Christensen}, {Morin}, {Gastine},
  {Reiners}, {Poppenhaeger}  \& {Wolk}}{{Yadav} et~al.}{2015}]{yadav15}
{Yadav} R.~K.,  {Christensen} U.~R.,  {Morin} J.,  {Gastine} T.,  {Reiners} A.,
   {Poppenhaeger} K.,   {Wolk} S.~J.,  2015, \mn@doi [ApJ]
  {10.1088/2041-8205/813/2/L31}, \href
  {http://cdsads.u-strasbg.fr/abs/2015ApJ...813L..31Y} {813, L31}

\bibitem[\protect\citeauthoryear{{Zacharias}, {Finch}, {Girard}, {Henden},
  {Bartlett}, {Monet}  \& {Zacharias}}{{Zacharias} et~al.}{2012}]{zacharias12}
{Zacharias} N.,  {Finch} C.~T.,  {Girard} T.~M.,  {Henden} A.,  {Bartlett}
  J.~L.,  {Monet} D.~G.,   {Zacharias} M.~I.,  2012, VizieR Online Data
  Catalog, \href {http://adsabs.harvard.edu/abs/2012yCat.1322....0Z} {1322}

\makeatother
\end{thebibliography}

%%%%%%%%%%%%%%%%%%%%%%%%%%%%%%%%%%%%%%%%%%%%%%%%%%

%%%%%%%%%%%%%%%%% APPENDICES %%%%%%%%%%%%%%%%%%%%%

%\appendix
%\section{Some extra material}
%If you want to present additional material which would interrupt the flow of the main paper, it can be placed in an Appendix which appears after the list of references.

%%%%%%%%%%%%%%%%%%%%%%%%%%%%%%%%%%%%%%%%%%%%%%%%%%

% Don't change these lines
\bsp	% typesetting comment
\label{lastpage}
\end{document}